\documentclass[12pt]{article}
\usepackage{amssymb,amsmath,bbold}
\usepackage{graphicx}
\usepackage{hyperref}
\numberwithin{equation}{section}

 \newcommand{\be}{\begin{eqnarray} } 
  \newcommand{\ee}{\end{eqnarray} } 

\begin{document}
\baselineskip 0.6cm

\def\simgt{\mathrel{\lower2.5pt\vbox{\lineskip=0pt\baselineskip=0pt
           \hbox{$>$}\hbox{$\sim$}}}}
\def\simlt{\mathrel{\lower2.5pt\vbox{\lineskip=0pt\baselineskip=0pt
           \hbox{$<$}\hbox{$\sim$}}}}

\begin{titlepage}

\begin{flushright}
\end{flushright}

\vskip 1.7cm

\begin{center}

{\Large \bf 
Yukawa Unification and the Superpartner Mass Scale
}

\vskip 0.8cm

{\large Gilly Elor, Lawrence J. Hall, David Pinner and Joshua T. Ruderman}

\vskip 0.4cm

{\it Berkeley Center for Theoretical Physics, Department of Physics, \\
     and Theoretical Physics Group, Lawrence Berkeley National Laboratory, \\
     University of California, Berkeley, CA 94720, USA} \\

\vskip2cm

\abstract{
Naturalness in supersymmetry (SUSY) is under siege by increasingly stringent LHC constraints, but natural electroweak symmetry breaking still remains the most powerful motivation for superpartner masses within experimental reach.  If naturalness is the wrong criterion then what determines the mass scale of the superpartners?  We motivate supersymmetry by (1) gauge coupling unification, (2) dark matter, and (3) precision $b-\tau$  Yukawa unification.  We show that for an LSP that is a bino-Higgsino admixture, these three requirements lead to an upper-bound on the stop and sbottom masses in the several TeV regime because the threshold correction to the bottom mass at the superpartner scale is required to have a particular size.  For $\tan \beta \approx 50$, which is needed for $t-b-\tau$ unification, the stops must be lighter than 2.8 TeV when $A_t$ has the opposite sign of the gluino mass, as is favored by renormalization group scaling.  For lower values of $\tan \beta$,  the top and bottom squarks must be even lighter.  Yukawa unification plus dark matter implies that superpartners are likely in reach of the LHC, after the upgrade to 14 (or 13) TeV, independent of any considerations of naturalness.  We present a model-independent, bottom-up analysis of the SUSY parameter space that is simultaneously consistent with Yukawa unification and the hint for $m_h = 125$~ GeV\@.  We study the flavor and dark matter phenomenology that accompanies this Yukawa unification.  A large portion of the parameter space predicts that the branching fraction for $B_s \rightarrow \mu^+ \mu^-$ will be observed to be significantly lower than the SM value. 
}

\end{center}
\end{titlepage}

\tableofcontents
\newpage
\section{Introduction and Conclusions}
\label{sec:intro}
For thirty years Weak Scale Supersymmetry has been a leading candidate for a deeper theory underlying the Standard Model (SM); it is motivated by a natural electroweak scale, precision unification of the gauge couplings, and Lightest Super Partner (LSP) dark matter from thermal freeze-out.  However, experiment has called into question the naturalness of the minimal theory, first by the LEP2 bound on the Higgs mass and second by direct superpartner searches and preliminary evidence for a Higgs boson with a mass near 125 GeV \cite{ATLAS:2012ae,Chatrchyan:2012tx} from 2011 LHC data.  If confirmed, the latter implies fine-tuning of the weak scale in the Minimal Supersymmetric Standard Model (MSSM) of at least 1\%~\cite{Hall:2011aa}.  Furthermore, LSP dark matter from thermal freeze-out requires a heavy gravitino, further increasing the fine-tuning.

Thus LHC data suggests two very different schemes for supersymmetry with precise gauge coupling unification: 
\begin{itemize}
\item  A natural weak scale.  An extension of the minimal model is required, for example by adding a gauge singlet superfield and/or violating R-parity.  LSP dark matter from thermal freeze-out may be lost or may require a considerable modification of the theory.
\item A fine-tuned weak scale.   The minimal model survives, allowing LSP dark matter from thermal freeze-out.
\end{itemize}
While both schemes were studied before the LHC, and while there are many avenues for supersymmetry with a natural weak scale, the possibility of a fine-tuned weak scale is growing in importance.

With a fine-tuned weak scale, the supersymmetry breaking scale $\tilde{m}$ becomes decoupled from the weak scale, so that discovery of superpartners at the LHC does not appear to be necessary.  For example, in Split Supersymmetry \cite{ArkaniHamed:2004fb} the scalar superpartners can have a mass $\tilde{m}$ many orders of magnitude larger than the weak scale, while only the fermionic superpartners are in the TeV domain.  Similarly in Anomaly Mediation \cite{Randall:1998uk,Giudice:1998xp} without sequestering, and in Spread Supersymmetry \cite{Hall:2011jd}, the scalar superpartner mass scale $\tilde{m}$ is larger than the weak scale by powers of an inverse loop factor (for early work along these lines see~\cite{Wells:2003tf, Wells:2004di}).  In all these theories supersymmetry may well be out of reach at the LHC.  

Are there any experimental constraints that limit $\tilde{m}$?  A Higgs mass near 125 GeV provides a constraint on $\tilde{m}$, but unfortunately only a weak one.  For example, in Split Supersymmetry $\tilde{m} < 10^5$ TeV \cite{Binger:2004nn,  Giudice:2011cg}.   More generally, at one extreme, with the Higgs quartic boosted by large top squark mixing, $\tilde{m} \sim$ TeV is possible.  At the other extreme, $\tilde{m} \sim 10^{13}$ TeV is possible if the Higgs quartic arises entirely from SM loop effects \cite{Hall:2009nd}, although this may require a threshold to the quartic coupling at the unified scale.  Flavor-changing and CP violating effects provide similarly weak constraints: if $\tilde{m} > 10^3$ TeV, virtual superpartners do not upset the successful CKM understanding of the SM even if the squark and slepton mass matrices have arbitrary off-diagonal entries.  However, this is clearly not a firm limit since the flavor matrices may have a hierarchical structure with small off-diagonal entries, as in the quark mass matrices.

In this paper we constrain $\tilde{m}$ by requiring precise unification of the $b$ and $\tau$ Yukawa couplings at the mass scale of gauge coupling unification~\cite{Georgi:1974sy, Chanowitz:1977ye, Buras:1977yy}.  Thus, our motivations for supersymmetry are the conventional ones, except naturalness is replaced by Yukawa unification:
\begin{itemize}
\item Precise unification of the gauge couplings.
\item Precise unification of the $b$ and $\tau$ Yukawa couplings.
\item Lightest Super Partner (LSP) dark matter from thermal freeze-out.
\end{itemize}

It seems significant that in the MSSM both gauge and Yukawa unification can result from very simple unified theories, for example, an $SO(10)$ theory with each generation in a 16-plet spinor, $\psi$, and the third generation masses arising from the single interaction $\psi_3 \psi_3 \phi$.  If $\phi$ unifies the two Higgs doublets of the MSSM, $H_{u,d}$, then  $t-b-\tau$ unification~\cite{Ananthanarayan:1991xp, Hall:1993gn} occurs, while if $\phi$ contains unequal components of $H_u$ and $H_d$ then $b-\tau$ unification results.   The smaller Yukawa couplings involving lighter generations may have a more complicated origin, for example from higher dimensional interactions, radiative effects, or environmental selection.

\begin{figure}[h!]
\begin{center} \includegraphics[width=0.75 \textwidth]{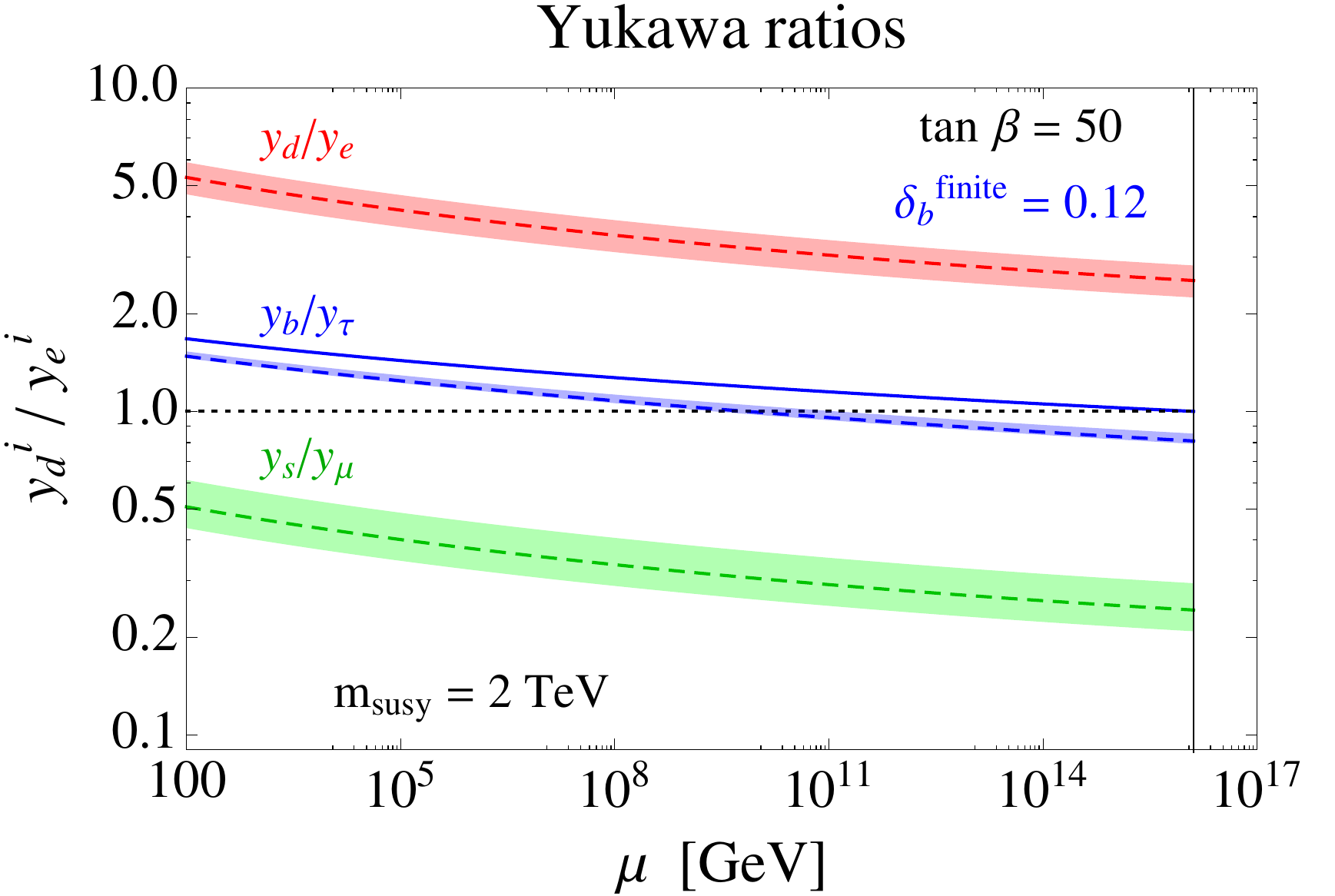}\end{center}
\caption{ \label{fig:YukawaRatio}
Yukawa couplings ratios for $d/e$, $s/\mu$ and $b/\tau$ as a function of scale, $\mu$.  The three dashed lines result from using the central values of the masses derived from experiment, two-loop running in the MSSM~\cite{Martin:1993zk} with all superpartners at 2 TeV and $\tan \beta = 50$, but ignoring the finite superpartner threshold corrections.  The colored bands allow for $\pm 2 \sigma$ uncertainties in the quark masses.  The solid curve includes a 12\% finite correction at the TeV scale to the $b$ Yukawa coupling.  The running $\overline{MS}$ masses evaluated at $m_Z$ are taken from Ref.~\cite{Xing:2011aa}.
}
\end{figure}

Ignoring supersymmetric threshold corrections, two-loop renormalization group scaling in the MSSM yields $b-\tau$ unification with a precision of about 20\% at $\tan \beta < 10$, and $t-b-\tau$ unification with a precision of about 10\% at $\tan \beta \simeq 50$.  Given that the quark and charged lepton masses vary over six orders of magnitude, this Yukawa unification is very striking, perhaps the best hint we have of any symmetry structure underlying quark and lepton masses, although evaluating its significance is not easy.   In figure~\ref{fig:YukawaRatio}, the $d/e$, $s/\mu$ and $b/\tau$ Yukawa ratios are shown as a function of scale in the MSSM, with superpartners at 2 TeV and $\tan \beta = 50$.   The dashed lines ignore the finite ({\it i.e.} non-log) supersymmetric threshold corrections, and the shaded bands include $\pm 2 \sigma$ uncertainties on the quark masses.   The $d/e$ and $s/\mu$ Yukawa ratios at the unified scale are close to the Georgi-Jarlskog  \cite{Georgi:1979df} values of 3 and 1/3, but in this paper we focus on the third generation.   The solid line shows the effect of including a 12\% finite supersymmetric threshold correction for the $b$ Yukawa coupling, as needed for precise Yukawa unification.      

We will conclude that the above three motivations for supersymmetry lead to $\tilde{m} \sim 1-10$~TeV\@.  Yet, once naturalness is ignored (or multiverse arguments are made for the weak scale and the dark matter abundance) one might question whether such a supersymmetric theory is preferred over the SM with axionic dark matter.  It is well-known that gauge coupling unification occurs more accurately in the supersymmetric case; here we stress that $b/\tau$ unification in the SM requires a unified threshold correction of 60\% -- a factor 5 larger than the supersymmetric threshold correction required for $t-b-\tau$ unification.  The combination of gauge and Yukawa unification is a powerful argument for (multi-) TeV scale supersymmetry. 

The key question becomes the origin of the 10 -- 20\% correction necessary for Yukawa unification.  Since unified threshold corrections are small for gauge couplings, $\leq 1$\%, they are also expected to be small for Yukawa couplings.  Although there is large SU(5) breaking in the $s/\mu$ Yukawa ratio, this typically affects the third generation $b/\tau$ boundary condition at the level of less than 1\%.   Thus, while it is possible to construct theories which have the required $b/\tau$ correction arising entirely from either boundary condition or threshold corrections at the unified scale, in a very wide class of theories the correction must arise from the supersymmetric threshold.   

A large 10 -- 20\% correction from the supersymmetric threshold could come from logarithmic terms, with at least some superpartners far above the weak scale, or it could come from finite threshold corrections.   With $\tilde{m}$ at the TeV scale,  figure~\ref{fig:YukawaRatio} shows that a finite threshold correction is needed.  We find that raising $\tilde{m}$ above the TeV scale does not eliminate the requirement of a finite threshold correction, whether all superpartners are raised together or Higgsinos and/or gauginos are kept at the TeV scale.  Furthermore, such large finite corrections are easy to obtain since they are enhanced by $\tan \beta$, the ratio of electroweak VEVs \cite{Hall:1993gn}.  In particular, the leading contribution to the finite bottom threshold in the heavy squark limit is 

\begin{figure}[t]
\begin{center} \includegraphics[width=\textwidth]{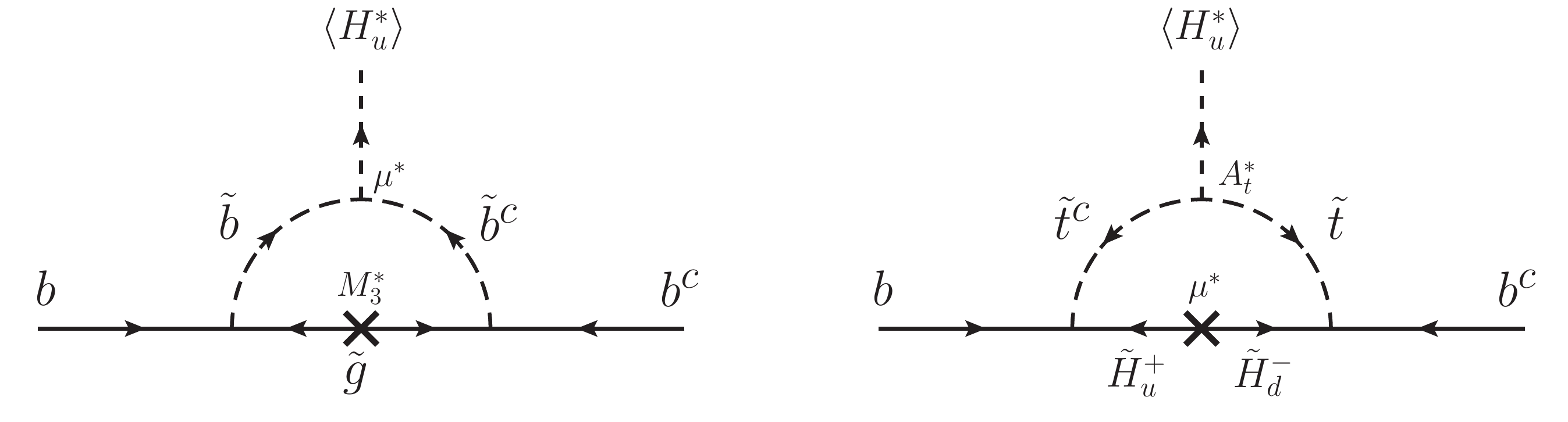}  \end{center}
\caption{
\label{fig:bthresh_feyn}
The leading, $\tan \beta$ enhanced contributions to the finite part of the bottom Yukawa threshold.
}
\end{figure}

\be \label{eq:bthresh}
\delta_b^{fin}=-\frac{g_3^2}{12\pi^2} \frac{\mu M_3}{m_{\tilde b}^2} \tan \beta - \frac{y_t^2}{32\pi^2} \frac{\mu A_t}{m_{\tilde t}^2} \tan \beta.
\ee
Two schemes allowing Yukawa unification at large $\tan \beta$ were introduced in Ref.~\cite{Hall:1993gn}.  For light squarks of a few hundred GeV individual contributions to $\delta_b^{fin}$ are typically too large, but there could be a cancellation.  With somewhat heavier squarks the contributions become suppressed, with $\delta_b^{fin} \propto \mu/ \tilde{m}$ where $\mu$ is the Higgsino mass parameter, and can give the desired 10-20\% correction without a cancellation.  Motivated by naturalness, many studies of Yukawa unification have focussed on the light squark case (for example Refs.~\cite{Rattazzi:1995gk, Baer:2000jj, Blazek:2001sb, Blazek:2002ta}); although Tobe and Wells were early advocates of easing the naturalness requirement, studying Yukawa unification with heavier squarks \cite{Tobe:2003bc}. 

We concentrate on a Higgsino/bino LSP in the MSSM, so that the dark matter abundance constrains the Higgsino mass parameter, $\mu \leq 1$ TeV\@. As the scale of supersymmetry breaking is increased above $\mu$, the finite supersymmetric threshold corrections to $b/\tau$ become suppressed, as they are proportional to $\mu/\tilde{m}$.  Since a large 10 -- 20\% finite supersymmetric threshold correction is necessary for exact Yukawa unification, there is an upper bound on $\tilde{m}$.  For $\tan \beta = 50$  we will see that this limits the third generation squark mass to be less than about 2.5~TeV when $A_t$ has the opposite sign of $M_3$, as is favored by Renormalization Group (RG) scaling\footnote{The gluino mass, $M_3$, appears in the RG for $A_t$ (see equation~\ref{eq:AtRG}) and drives it to have the opposite sign unless $A_t$ is chosen to have a large value in the UV.}.  As $\tan \beta$ is reduced this limit becomes stronger.  Thus the simplest supersymmetric scheme for Yukawa unification and LSP dark matter imposes a powerful constraint on $\tilde{m}$, even when naturalness is ignored.  When $A_t$ and $M_3$ have the same sign, the limit weakens to about 9~TeV for $\tan \beta= 50$. 

Furthermore, with $\tilde{m} \lesssim 1$~TeV a variety of constraints make the theory implausible.   Yukawa unification with large $\tan \beta$ leads to excessive contributions to $b \rightarrow s \gamma$ and to $\delta_b^{fin}$, requiring cancellations in both cases.  A Higgs mass near 125 GeV is hard to achieve, and LHC bounds on squark production become stringent.   Hence we are led to Yukawa unification with squark masses in the several TeV range.

Since we are dropping naturalness, one might wonder if there is a supersymmetric spectrum having squark masses many orders of magnitude above the weak scale that has both gauge and Yukawa unification with small unified thresholds and LSP dark matter.  In split supersymmetry this is not the case, as the additional logarithmic supersymmetric threshold corrections to $b-\tau$ unification have the wrong sign, while the finite corrections essentially vanish~\cite{Giudice:2004tc}.   Having only the Higgsinos at the weak scale leads to the same situation, while keeping only the gauginos at the weak scale significantly degrades gauge coupling unification.

The organization of this paper is as follows.  In section~\ref{sec:ModIndep}, we perform a model independent analysis of $b-\tau$ unification.  As a function of the supersymmetric threshold corrections to the $t$, $b$, and gauge couplings,  we find the parameter choices that give precision Yukawa unification with a small GUT-scale threshold correction.
 In section~\ref{sec:unifiedscale} we discuss various aspects of the GUT-scale boundary condition and threshold corrections.
 We begin in subsection~\ref{subsec:unifiedbc} by showing that it is possible for the first two generations to have non-unified boundary conditions, and to have the necessary CKM mixing with the third generation, without generating large corrections to $b/\tau$.  Typically GUT threshold corrections are small, but in subsection~\ref{subsec:unifiedthresh} we discuss two possible sources of large corrections: (1) a large value of $y_t$ near the GUT-scale (which results from low $\tan \beta$) and, (2) the threshold corrections that arise from Yukawa unification in an orbifold GUT\@.
 
 Our detailed exploration of the MSSM parameter space begins in section~\ref{sec:susy_tbt}, where we determine which parameters yield both precision $b-\tau$ unification and a 125 GeV Higgs mass.  By imposing that bino-Higgsino dark matter does not overclose, $\mu \le 1$~TeV, we arrive  at upper bounds on the stop and sbottom masses in the several TeV regime. Having identified the allowed parameter space, we move on to study its phenomenology in section~\ref{sec:signals}.    In subsection~\ref{subsec:Bmeson}, we consider SUSY contributions to the rare $B$-meson decays, $b \rightarrow s \gamma$ and $B_s \rightarrow \mu^+ \mu^-$.  For the parameters that give Yukawa unification, $b \rightarrow s \gamma$ is an important constraint, but a large portion of the parameter space is allowed without requiring fine-tuning.  Meanwhile, we find that $B_s \rightarrow \mu^+ \mu^-$ is observably depleted  in much of the parameter space, and therefore provides a promising discovery handle.   We present the dark matter phenomenology in subsection~\ref{subsec:dm}.  We determine the relic abundance and direct detection cross-section within the parameter space that gives $b-\tau$ unification.  Although the present limits from direct detection are weak, we find that XENON1T will probe a sizable fraction of the parameter space.  In appendix~\ref{app:ExtraCharged} we consider how extra charged states modify Yukawa unification.  In appendix~\ref{app:flavor}, we describe our notation for various flavor bases that are useful for understanding rare $B$-meson decays.  In appendix~\ref{app:welltemp} we present an updated, model-independent, study of the parameter space of the well-tempered neutralino~\cite{ArkaniHamed:2006mb}, including recent lattice values for the strange quark content of the nucleon~\cite{Giedt:2009mr}.

\section{Model-Independent Analysis of Threshold Corrections}
\label{sec:ModIndep}

In any scheme that embeds the SM in a supersymmetric theory at scale $\tilde{m}$ and has quark-lepton and gauge coupling unification at scale $M_U$, the couplings will receive threshold corrections at both the supersymmetric and unified scales.  In this section we present a general analysis of the necessary size of the threshold corrections, without appealing to specific supersymmetric spectra.  Armed with the model-independent results of this section, we will consider explicit spectra  in section~\ref{sec:susy_tbt}.

   We match the gauge couplings, $g_a$, with $a=3,2,1$ for $SU(3), SU(2), U(1)$ and the third generation Yukawa couplings, $y_i$ with $i=t,b,\tau$ of the SM to those of the MSSM by introducing threshold corrections $\delta_{i,a}$
\begin{equation}
y_t^{MSSM}(M_Z) = \frac{y_t^{SM}(M_Z)}{ \sin \beta} \, (1+\delta_t),  \hspace{1in} y_{b,\tau}^{MSSM}(M_Z) = \frac{y_{b,\tau}^{SM}(M_Z)}{\cos \beta} \, (1+\delta_{b,\tau})
\label{eq:deltai}
\end{equation}
\begin{equation}
g_a^{MSSM}(M_Z) = g_a^{SM}(M_Z) (1+\delta_a). 
\label{eq:deltaa}
\end{equation}
Although we are interested in $\tilde{m}$ up to three orders of magnitude above the weak scale, for simplicity  we define the above thresholds at the reference scale $M_Z$.   The unified couplings result from matching to the MSSM couplings at $M_U$
\begin{equation}
y(M_U) = y_b^{MSSM}(M_U) (1+\epsilon) = y_\tau^{MSSM}(M_U)
\label{eq:epsilon}
\end{equation}
and 
\begin{equation}
\frac{1}{{\bar g}^2(M_U)} =\frac{1}{g_a^2(M_U)} (1+\epsilon_a) 
\label{eq:epsilona}
\end{equation}
where $1/{\bar g}^2$ is the average of $1/g_a^2$.   The unified scale, $M_U$, is defined by minimizing $\epsilon_g^2 = \sum_a \epsilon_a^2$.  Between the matching scales we evolve the couplings using two-loop MSSM renormalization group equations~\cite{Martin:1993zk}, so that $b-\tau$ unification leads to a constraint between the thresholds
\begin{equation}
\epsilon = \epsilon(\delta_{i,a}, \tan \beta). 
\label{eq:btauconstraint}
\end{equation}
When specializing to $t-b-\tau$ unification, $\tan \beta$ is fixed by setting $y_t=y_b$ at $M_U$.

The SUSY threshold corrections, $\delta_{i,a}$, depend on the superpartner spectrum and at one-loop are the sum of terms proportional to the log of superpartner masses, $\delta_{i,a}^{log}$, and terms that are independent of logarithms of superpartner masses, $\delta_{i,a}^{fin}$.  Setting all superpartner masses equal, gauge coupling unification is very precise, with $\epsilon_g = 0.013 (0.017)$ for $\tilde{m} = 0.1 (100)$~TeV\@.  For superpartner masses in the range of interest to us, gauge coupling unification is remarkably insensitive to the supersymmetric thresholds.  On the other hand, for the example shown in figure~\ref{fig:YukawaRatio}, with all superpartners at $\tilde{m} = 2$ TeV, $\tan \beta = 50$, and $\delta_{t,b,\tau,a}^{fin} $ all set to zero, $b-\tau$ unification requires large unified thresholds, $\epsilon = 0.23$. We now analyze more generally the constraint on the thresholds, (\ref{eq:btauconstraint}), required for $b-\tau$ and $t-b-\tau$ unification.

\begin{figure}[h!]
\begin{center} \includegraphics[width=\textwidth]{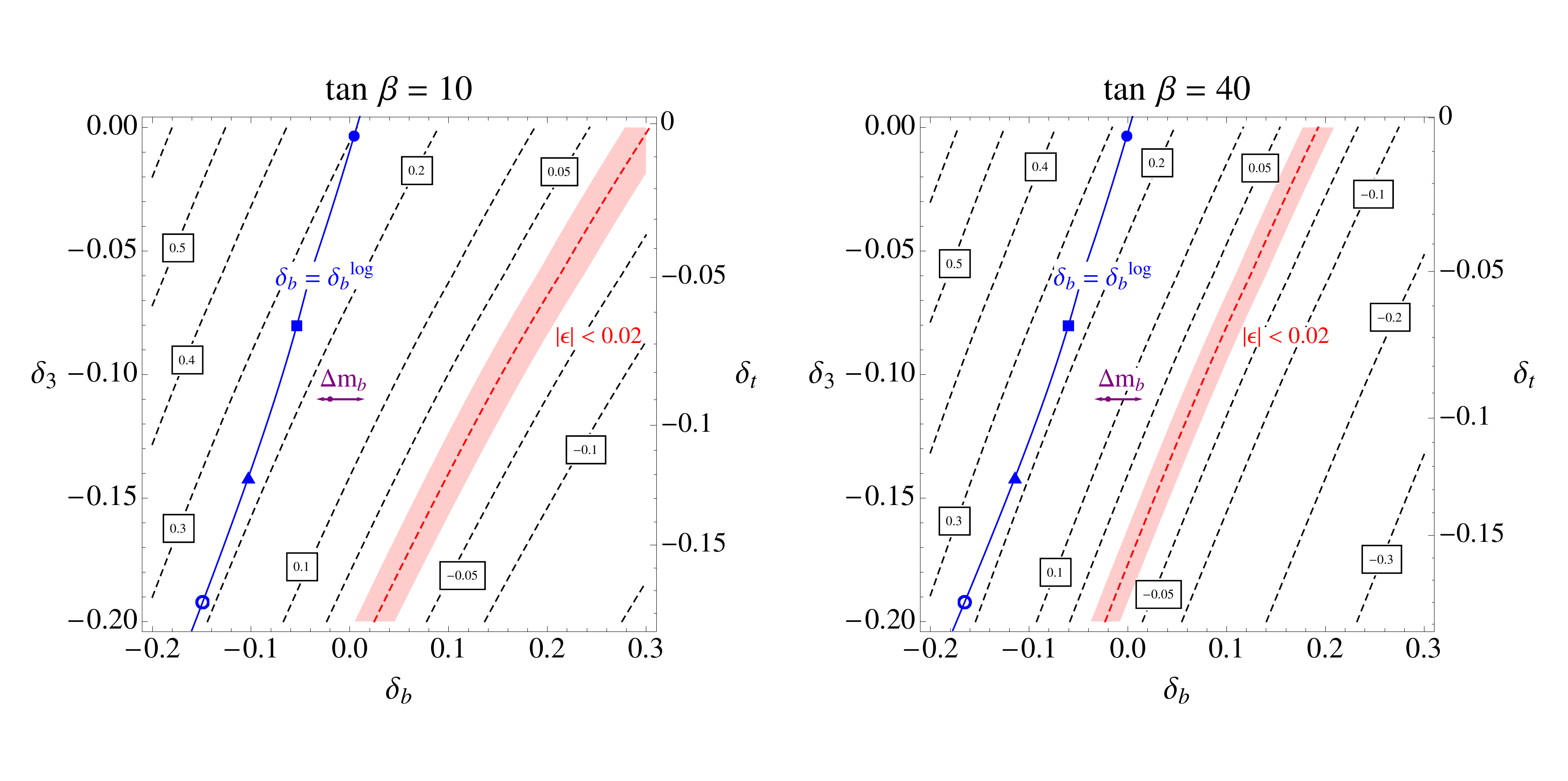} \end{center}
\caption{ \label{fig:ModelIndep_btau}
Contours of the GUT-scale threshold $\epsilon$, as a function of the $y_b$ and $g_3$ supersymmetric thresholds, for $\tan \beta = 10, 40$.  We define $\epsilon$ by $y_b(1+\epsilon) = y_\tau$ at the GUT scale, and the red region with $|\epsilon| < 0.02$ has precision Yukawa unification.  The blue contour shows the logarithmic contribution to $\delta_b$ for a particular spectrum with degenerate scalar masses, described in the text. The purple arrow represents the $2 \sigma$ experimental uncertainty on the bottom mass.  The top Yukawa threshold, $\delta_t$, is chosen to vary with $\delta_3$, also assuming a spectrum with degenerate scalars.  For simplicity, we have set $\delta_1 = \delta_2 = 0$ since these threshold corrections are numerically unimportant for Yukawa unification, and we have set $\delta_\tau = 0$ since this threshold is typically very small.
}
\end{figure}

In practice the threshold $\delta_{\tau}$ is small enough to be neglected, and $\delta_{1,2}$ have a small impact on the $ b /\tau$ ratio, relative to $\delta_3$.  Therefore, for simplicity, we set $\delta_{\tau, 1, 2} = 0$ in this section, and we study $\epsilon(\delta_{t,b,3}, \tan \beta)$.  Furthermore, the dependence of $\epsilon$ on $\delta_t$ is significantly weaker than on $\delta_{b,3}$.   Hence we use an approximation for $\delta_t$, obtained by taking all superpartners degenerate except the Higgsinos, which have mass $\mu$ = 500 GeV, and ignoring finite threshold corrections to $y_t$:  $\delta_t^{approx} = \delta_t^{approx}(\delta_3)$.  Below, we will show that deviating from this approximation for $\delta_t$ does not qualitatively change our results.   Using this approximation we can draw contours of $\epsilon$ in the $\delta_b - \delta_3$ plane as shown by the dashed lines in figure~\ref{fig:ModelIndep_btau}, with $\tan \beta = 10 (40)$ in the left (right) panel.   These dashed lines are model independent; for any values of $\delta_{3,b}$ they yield the size of the unified threshold correction necessary for $b-\tau$ unification.   The dashed red line is the $\epsilon=0$ contour that gives precise $b-\tau$ unification without any unified thresholds, while the red shaded area corresponds to unified threshold corrections with magnitude less than 2\%. The running $\overline{MS}$ value of the $b$ quark mass is taken to be $m_b(m_b) = (4.19+0.18-0.06)$~GeV at $3 \sigma$~\cite{pdg}; the purple line in figure~\ref{fig:ModelIndep_btau} shows the $2 \sigma$ experimental uncertainty in $m_b$.  We determine the $\overline{DR}$ bottom mass, evaluated at $M_Z$, using the 2-loop conversion from $\overline{MS}$~\cite{Baer:2002ek} and a 4-loop approximation to running $\alpha_s$~\cite{Bethke:2009jm}.
  
\begin{figure}[h!]
\begin{center} \includegraphics[width=\textwidth]{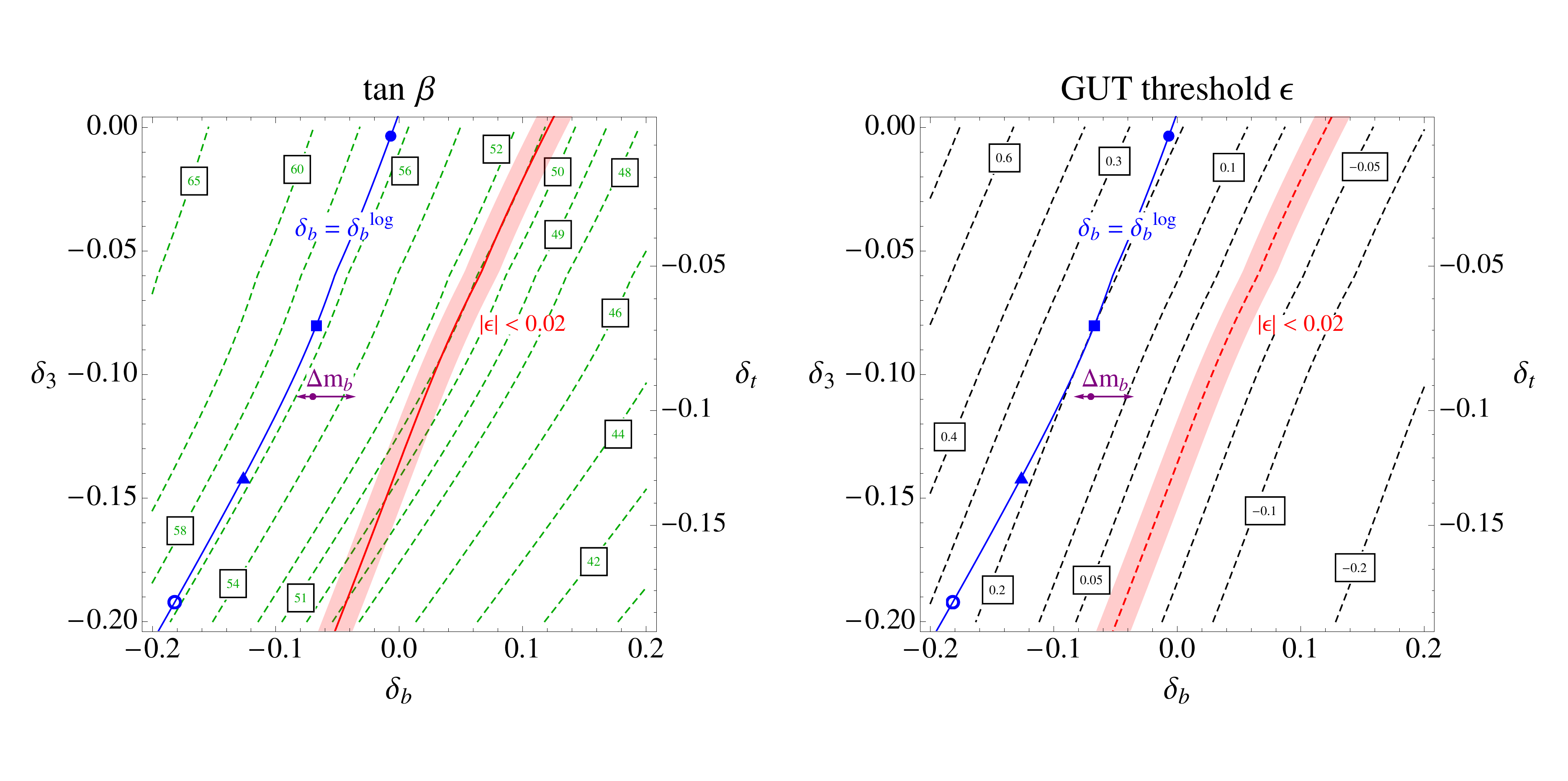}  \end{center}
\caption{ \label{fig:ModelIndep_tbtau}
The necessary value of $\tan \beta$ for $t-b-\tau$ unification ({\it left}), and the GUT-scale Yukawa threshold, $\epsilon$ ({\it right}), as a function of the $y_b$ and $g_3$ thresholds.  At each point, $\tan \beta$ (shown as green contours to the left) is chosen so that $y_t = y_b$ at the scale of gauge coupling unification.  On the right, the GUT threshold, $\epsilon$, is defined by the relation $y_b (1+\epsilon) = y_\tau$.  As in figure~\ref{fig:ModelIndep_btau}, we set the $g_{1,2}$ and $y_\tau$ thresholds to 0 and $\delta_t$ is varied with $\delta_3$ assuming degenerate scalar superpartners.
}
\end{figure}

The blue contour in figure~\ref{fig:ModelIndep_btau} shows the relationship between $\delta_3$ and $\delta_b$ that results from ignoring all finite thresholds and using the supersymmetric spectrum, described above, with degenerate scalars that was used to determine $\delta_t^{approx} (\delta_3)$.   The superpartner mass scale $\tilde{m}$ increases from top to bottom along the blue contour, taking values of (0.1,1,10,100) TeV at the (circle, square, triangle, open circle).  Note that the finite corrections to $\delta_{t}^{fin}$ are negligible since they are not $\tan \beta$ enhanced, and there are no finite corrections to $g_3$ in the $\overline{DR}$ scheme~\cite{Pierce:1996zz}.  The absence of an intersection between the red and blue lines throughout this range of $\tilde{m}$ and $\tan \beta$ implies that Yukawa unification requires $\epsilon$ and/or $\delta_b^{fin}$ to be non-zero.
For example, $\delta_b^{fin}=0$ requires rather large $\epsilon = $ 0.2 to 0.3.   Alternatively, for a given $\tilde{m}$ and $\tan \beta$, one can use this figure to determine the required value of $\delta_b^{fin}$ to yield $\epsilon=0$ by reading off the horizontal distance between the blue and red contours.  As $\tan \beta$ decreases, larger values of $\delta_b^{fin}$ are required\footnote{Larger $\delta_b^{fin}$ is required at lower values of $\tan \beta$ because for larger $\tan \beta$, $y_b$ is larger and its self-running increases $y_b / y_\tau$. }, and since $\delta_b^{fin} \propto \tan \beta$ in this region of $\tan \beta$, there is a lower limit on $\tan \beta$ that allows $\epsilon=0$.  When we consider explicit spectra in section~\ref{sec:susy_tbt}, we will find that $\tan \beta > 10$ is necessary when the stops and sbottoms are degenerate.

\begin{figure}[h!]
\begin{center} \includegraphics[width=\textwidth]{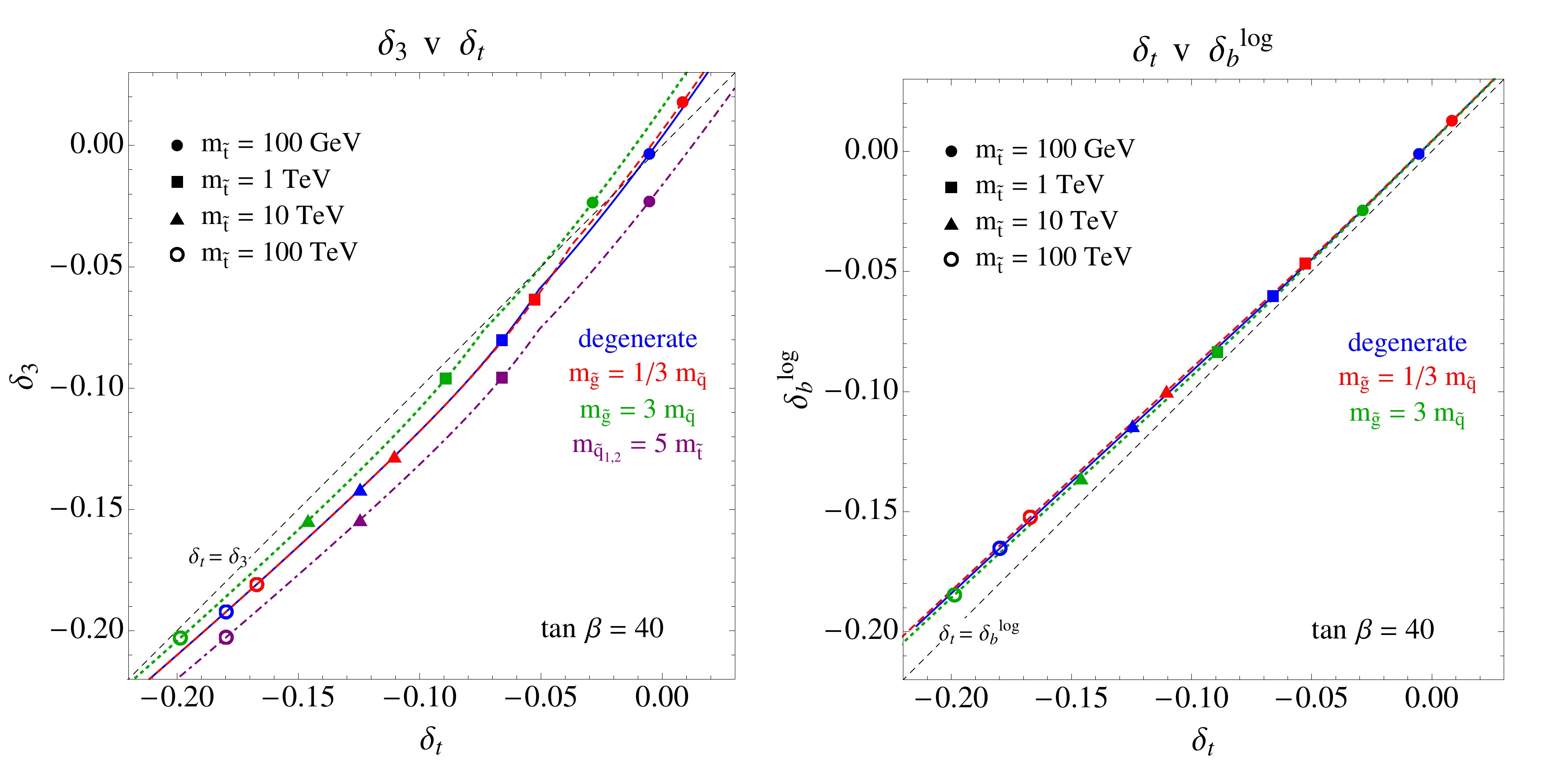}\end{center}
\caption{ \label{fig:LogThresh}
The relationship between $\delta_t$ and $\delta_3$ ($\delta_t$ and $\delta_b^{log}$)  is shown to the {\it left} ({\it right}) for various  spectra.  For all spectra we fix $\mu = 500$~GeV, assume gaugino unification, and use markers to indicate stop masses of $m_{\tilde t} = 0.1, 1, 10, 100$~TeV\@.   For the blue spectrum, the scalars and gluino are degenerate with mass $\tilde m$.  For the red (green) spectrum, $M_3 = 1/3 \, (3) \, m_{\tilde q}$.   For the purple spectrum, $M_3 = m_{\tilde t,b}$, and the first and second generation squarks are 5 times heavier than the stop and sbottom.  In both plots, the variation among spectra is remarkably mild, justifying our simplifying assumption (the blue spectrum) used to relate $\delta_t$ and $\delta_b^{log}$ to $\delta_3$ in figures~\ref{fig:ModelIndep_btau} and~\ref{fig:ModelIndep_tbtau}. 
}
\end{figure}

This analysis is extended to the case of $t-b-\tau$ unification in figure~\ref{fig:ModelIndep_tbtau}.  In the left panel, for each value of $(\delta_b,\delta_3)$ a value of $\tan \beta$ is determined by requiring $y_t=y_b$ at $M_U$, yielding the green dashed contour lines.  In the right panel the dashed lines give the resulting contours for $\epsilon$.  The $\epsilon=0$ contour is highlighted in red in both panels, with red shading denoting the region with $| \epsilon | < 0.02$. In both panels the blue contour shows the relationship between $\delta_3$ and $\delta_b$ that results from ignoring all finite thresholds and using the same supersymmetric spectrum as in figure~\ref{fig:ModelIndep_btau}.   As $\tilde{m}$ is increased from 0.1 to 100 TeV, $\epsilon=0$ requires values of $\tan \beta$ from 51 to 48. For $\delta_b^{fin}=0$, $\tan \beta$ is near 56.  As with lower values of $\tan \beta$, sizable values of $\epsilon$ and/or $\delta_b^{fin}$ are required.

We conclude this section by testing the robustness of some of the assumptions that we made in order to construct figures~\ref{fig:ModelIndep_btau} and~\ref{fig:ModelIndep_tbtau}.  In particular, we chose a specific spectrum in order to relate $\delta_t$ to $\delta_3$ (which are shown on the right and left vertical axes, respectively) and in order to draw the blue curve that relates $\delta_b^{log}$ to $\delta_3$.  The left-side of figure~\ref{fig:LogThresh} shows $\delta_t(\delta_3)$ for alternative spectra having the gluino mass larger or smaller than the degenerate squarks by a factor of three, or having the squarks of the first two generations a factor of five heavier than those of the third generation.  The variations among these spectral lines are remarkably mild, justifying our approximate choice of $\delta_t (\delta_3)$ in figures~\ref{fig:ModelIndep_btau} and~\ref{fig:ModelIndep_tbtau}.  On the right of figure~\ref{fig:LogThresh}, we show $\delta_t$ versus $\delta_b^{log}$ for light and heavy gluino, and we see that there is no variation in the functional dependence for different spectra (the only thing that changes somewhat is the relationship between the scalar mass and the position in the $\delta_t-\delta_b$ plane).  Since $\delta_b^{log}$ is a fixed function of $\delta_t$, and $\delta_t$ is a nearly fixed function of $\delta_3$, these plots together imply that $\delta_b^{log}$ is a roughly fixed function of $\delta_3$.  This means that the position of the blue curve in figures~\ref{fig:ModelIndep_btau} and~\ref{fig:ModelIndep_tbtau}, defined by $\delta_b = \delta_b^{log}$, also shows only mild model-dependence.

\section{Corrections from the Unified Scale}
\label{sec:unifiedscale}

Over a wide range of parameters, a 10--20\% adjustment to $y_b/y_\tau$ is required, as shown in figures~\ref{fig:ModelIndep_btau} and~\ref{fig:ModelIndep_tbtau}.   A key question is how much of this correction originates from the finite contribution to the supersymmetric threshold, $\delta_b^{fin}$, and how much from the unified scale, $\epsilon$.  If the entire correction is accounted for by $\epsilon$, then the combination of $b-\tau$ unification and WIMP LSP dark matter does not constrain the scale $\tilde{m}$ of superpartner masses.

The unified correction can arise from a tree-level breaking of the SU(5) boundary condition, or from loop threshold corrections that depend on the spectrum of the states at the unified scale
\begin{equation}
\epsilon \, = \,  \epsilon_{b.c.} \, + \, \epsilon_{th}
\label{eq:epsilon2}
\end{equation}
and we consider these in turn.

\subsection{Boundary Condition Corrections}
\label{subsec:unifiedbc}

We focus on unified theories where the dominant contribution to the $b$ and $\tau$ Yukawa couplings are equal, arising from an operator that preserves SU(5).    In SU(5) notation, this results from the operator $T_3 \bar{F}_3 \bar{H}$, where $T_i, \bar{F}_i$ are the 10 and $\bar{5}$ matter fields for the three generations $i=1,2,3$.   
We allow SU(5) violation in interactions involving $(T_2 \bar{F}_3,  T_3 \bar{F}_2,  T_2 \bar{F}_2)$ that lead to the CKM element $V_{cb}$ and the mass ratios $m_s/m_b$, $m_\mu/m_\tau$.  From figure~\ref{fig:YukawaRatio}, SU(5) violation in these mass ratios is required.   Given the comparable magnitude of these three quantities -- they are all within a factor of 2 of 1/30 -- we also allow SU(5) breaking in the mixing between the two heavy generations.   At what level does this SU(5) breaking feed into the $b-\tau$ boundary condition?

With the above assumptions, the Yukawa coupling matrices for the heaviest two generations in the down (D) and charged lepton (E) sectors of the form
\begin{equation}
y_{D,E} = 
\left( \begin{array}{cc}
 B_{D,E} \, \delta &C_{D,E} \,  \delta \\
A'_{D,E} & 1 \end{array} \right) \, y
\label{eq:yDE1}
\end{equation}
where $ B_{D,E},  C_{D,E}$ are order unity and $\delta = 1/30$ governs the scale of  $(V_{cb}, m_s/m_b, m_\mu/m_\tau)$.   Theories with flavor symmetries acting on $T_i$ but not $\bar{F}_i$ lead to neutrino anarchy, and to hierarchies in the up sector that are roughly the square of those in the (D,E) sectors; furthermore, they have $A'_{D,E}$ of order unity.  This leads to large 23 mixing angles on the right-handed quarks and left-handed leptons, $\tan \theta_{D,E} = A'_{D,E}$, giving a large correction to the boundary condition: $y_b/y_\tau = (\cos \theta_E/\cos \theta_D) (1 + {\cal O} (\delta^2))$.   Hence we assume that any contribution to $A'_{D,E}$ larger than ${\cal O} (\delta)$ is SU(5) invariant
\begin{equation}
A'_{D,E} = A + A_{D,E} \, \delta,
\label{eq:ADE}
\end{equation}
where $A_{D,E} \sim {\cal O} (1)$.  

After rotating the right-handed states by $\theta$, with $\tan \theta = A$, and redefining parameters, the Yukawa matrices take the form
\begin{equation}
y_{D,E} = 
\left( \begin{array}{cc}
 B_{D,E} \, \delta &C_{D,E} \,  \delta \\
A_{D,E} \, \delta & 1 \end{array} \right) \, y.
\label{eq:yDE2}
\end{equation}
Many theories, especially those based on SO(10), have $A=0$ and give this form directly.  This form
is diagonalized by small ${\cal O}(\delta)$ angles giving
\begin{equation}
\epsilon_{b.c.} =  \frac{\delta^2}{2} \; (C_E^2 - C_D^2 + A_E^2 - A_D^2) \, \sim \, 10^{-2} - 10^{-3}.
\label{eq:epsbc}
\end{equation}
Hence, in theories of hierarchical charged fermion masses the simplest expectation is $\epsilon_{b.c.} \sim \delta^2 \sim V_{cb}^2 \sim 10^{-3}$.  However, to obtain $m_s/m_\mu \sim 3$ at the unified scale requires a ratio of Clebsch factors of 3 \cite{Georgi:1979df}, giving the possibility $\epsilon_{b.c.} \sim 3^2 \delta^2 \sim (m_\mu/m_\tau)^2 \sim 10^{-2}$; hence the range given in equation~(\ref{eq:epsbc}).  Such a Clebsch might originate in the 22 entry of the Yukawa matrix and be transferred to $C_{D,E}$ via a large 23 rotation due to $A \sim {\cal O} (1)$, or it might be present in the original 23 entry.

A wide class of unified theories, having Yukawa matrices of (\ref{eq:yDE2}), lead to $\epsilon_{b.c.} <0.01$.  In these theories $b-\tau$ unification requires (10--20)\% threshold corrections from either supersymmetric or unified scales.

It is important to note that theories having an SU(5) breaking contribution to the 33 entry of $y_{D,E}$ of order $\delta$ will typically give $\epsilon_{b.c.} \sim \delta \sim$ 3\%.  If this is further enhanced it could allow $b-\tau$ unification without threshold corrections~\cite{Barr:2002mw}.  Consider an SO(10) theory where the Yukawa interactions involving the third generation are
\begin{equation}
y \, \psi_3 \psi_3 \, \phi \, + \, y' (\cos \theta \, \psi_3 + \sin \theta \, \psi_2)^2 \, \phi',
\label{eq:SO10}
\end{equation}
where the 16-plet spinors $\psi_i$ contain $T_i$ and $\bar{F}_i$.  The multiplets $\phi$ and $\phi'$ are 10 and 126 dimensional and contain components of the MSSM Higgs doublets $\phi \supset x_a H_a$, $\phi' \supset x_a' H_a$, $a=u,d$. A weak triplet in $\phi'$ acquires a vev leading to type II seesaw neutrino masses, with $\theta \sim 45^o$ the atmospheric neutrino mixing angle. The coupling $y'$ must be chosen such that the weak doublet vev in $\phi'$ yields $V_{cb}$, implying that the $ \psi_3 \psi_3 \phi'$ interaction leads to 
\begin{equation}
\epsilon_{b.c.} = 4 V_{cb} \, \frac{x}{1-x} \sim 0.1 \, \frac{x}{1-x}
\label{eq:epsbc2}
\end{equation}
where $x= x_u x_d'/ x_d x_u'$.

\subsection{Threshold Corrections}
\label{subsec:unifiedthresh}

Contributions to the unified threshold correction $\epsilon_{th}$ are typically comparable to those that correct the gauge couplings, $\epsilon_a$, since they both arise at 1-loop, involve the same couplings and the same logarithms of mass ratios.  The success of precision gauge coupling unification then implies that $|\epsilon_{th}| < 0.02$, so that it can be neglected.   However, there are exceptions to this situation and we consider two in this sub-section.  Some couplings contribute at 1-loop to logarithmic Yukawa thresholds but not gauge thresholds.  The top Yukawa coupling is an example, but is usually not large enough to give significant contributions to $\epsilon_{th}$.  However, if $\tan \beta$ is sufficiently reduced these corrections can become important.  Secondly, in orbifold grand unified theories the KK towers at the unified scale do make substantial corrections to gauge coupling unification, so it is important to understand how they may affect Yukawa unification.

\subsubsection{From the Top Coupling}
\label{subsubsec:top}

\begin{figure}[h!]
\begin{center} \includegraphics[width=\textwidth]{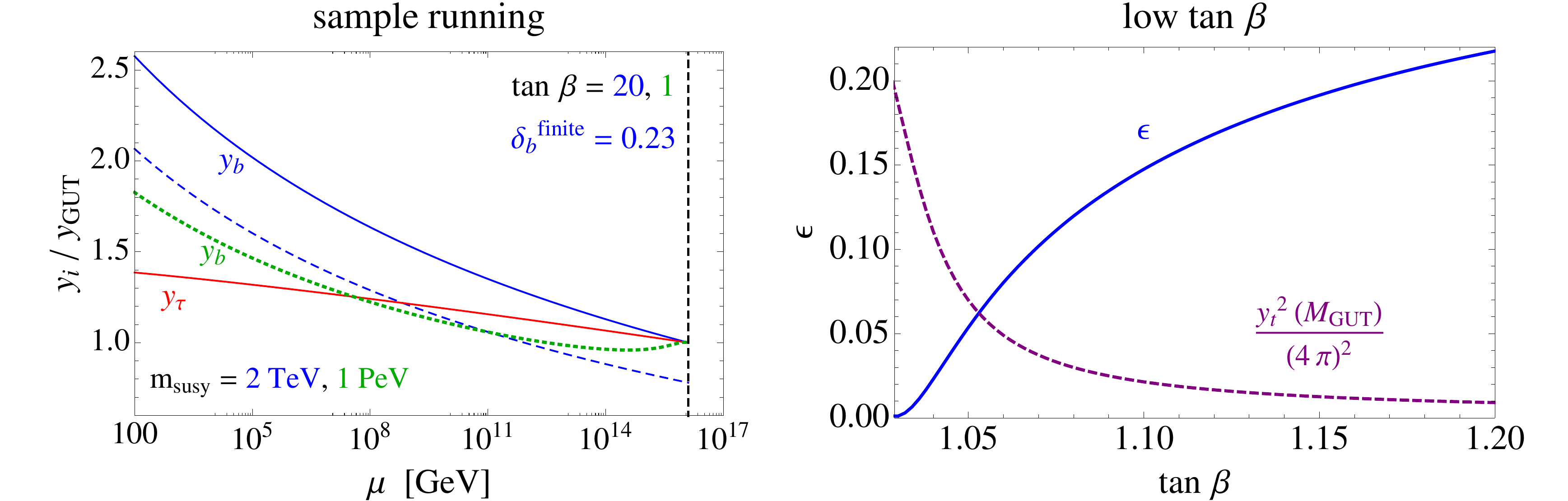}\end{center}
\caption{ \label{fig:lowtanbeta}  The left panel shows the evolution of the $b$ Yukawa coupling in two schemes.  The blue curves are for $\tan \beta = 20$ and squarks at 2 TeV, without (dashed) and with (solid) finite supersymmetric threshold corrections.  The green dotted curve has a split spectrum with squarks at $10^3$ TeV and $\tan \beta \approx 1$.  The red solid curve shows the evolution of the $\tau$ Yukawa coupling in both schemes.   In each scheme the curves are normalized by the value of the unified Yukawa coupling.  In the right panel the solid blue curve shows the required $\epsilon$ for Yukawa unification in the scheme with the split spectrum as $\tan \beta$ is varied, and the dashed purple line shows that the unified top Yukawa is rapidly increasing at low $\tan \beta$.
}
\end{figure}

At low $\tan \beta$ the top Yukawa coupling increases, changing the evolution of the $b$ quark Yukawa coupling in the direction of improving $b - \tau$ unification.  This can allow the squark mass scale to increase well beyond 10 TeV while keeping Higgsinos at the TeV scale for dark matter.  An example of this situation is shown in the left panel of figure~\ref{fig:lowtanbeta} where it is compared to the case of precision Yukawa unification at high $\tan \beta$.  The blue curves show the evolution of $y_b$ with $\tan \beta = 20$ for $\delta_b^{fin} =0$ (dashed) and 0.23 (solid).  The latter case leads to precision unification with the $\tau$ coupling, shown by the red solid curve.  The squark masses are taken to be 2 TeV and this is an illustration of the precision Yukawa unification discussed throughout this paper.  

The dotted green line illustrates the evolution of the $b$ Yukawa coupling for an entirely different possibility: split supersymmetry with squarks at $10^3$ TeV\@.  The fermionic superpartners are at the TeV scale with $\mu = 0.5$ TeV, $M_3 = 2$ TeV and gaugino mass unification.  Since the $A$ parameter breaks $R$ symmetry it is expected to be of order the gaugino masses and makes no relevant radiative contribution to the Higgs mass.  Indeed, we take $\tan \beta \approx 1$ so that the entire Higgs mass arises from RG scaling of the Higgs quartic between the top squark mass and the weak scale, requiring a stop quark mass of about $10^3$ TeV for a 125 GeV Higgs mass.  In this example all squark masses are set at $10^3$ TeV\@.  With such heavy squarks there is an important logarithmic supersymmetric threshold to $b/\tau$,  which works against Yukawa unification \cite{Giudice:2004tc}.  Hence the need for very low $\tan \beta$ leading to a large $y_t$.  

The effect of increasing $y_t$ is seen by comparing the dashed blue and green curves.   The difference in the running of these curves is most pronounced near the unification scale, where the top Yukawa coupling in the $\tan \beta = 1$ case is getting very large.   The dashed purple curve in the right panel of figure~\ref{fig:lowtanbeta} shows $y_t^2/16 \pi^2$ at the unification scale; it is less than 0.01 at large $\tan \beta$ and rises very rapidly to over 0.2 by $\tan \beta = 1.03$.  The corresponding reduction in the size of the required correction $\epsilon$ at the unified scale is shown by the solid blue curve.  Indeed at $\tan \beta = 1.03$ no correction is needed.  However, this cannot be viewed as precision Yukawa unification since unknown threshold corrections of order $y_t^2/16 \pi^2 = 0.2$ are expected at this point, even if they are not logarithmically enhanced.  Perhaps the best one can do is near $\tan \beta = 1.05$, where $\epsilon = 0.06$ is required and corrections of order $y_t^2/16 \pi^2 = 0.06$ are expected.   More generally the situation is clear:  Yukawa unification at low $\tan \beta$ does not have a precision that is controlled by the low energy theory, rather the required 10--20\% correction is coming from the unified scale where the top Yukawa coupling is approaching its Landau pole.

\subsubsection{From KK Towers in Orbifold GUTs}
\label{subsubsec:orbifoldGUTs}
Precision gauge coupling unification can occur in theories where the unified symmetry is realized in higher dimensions but is broken by the compactification to 4d.  The higher dimensional symmetry forbids Yukawa couplings from appearing in the bulk, and hence Yukawa unification can occur in such theories if the operator generating the $b$ and $\tau$ Yukawa couplings is located on a fixed point that respects SU(5).   Such Yukawa couplings do not have threshold corrections from KK modes because of locality.  However, gauge couplings do receive corrections from KK towers, and this can alter the scale of gauge coupling, and therefore Yukawa coupling, unification.   

\begin{figure}[h!]
\begin{center} \includegraphics[width=0.5\textwidth]{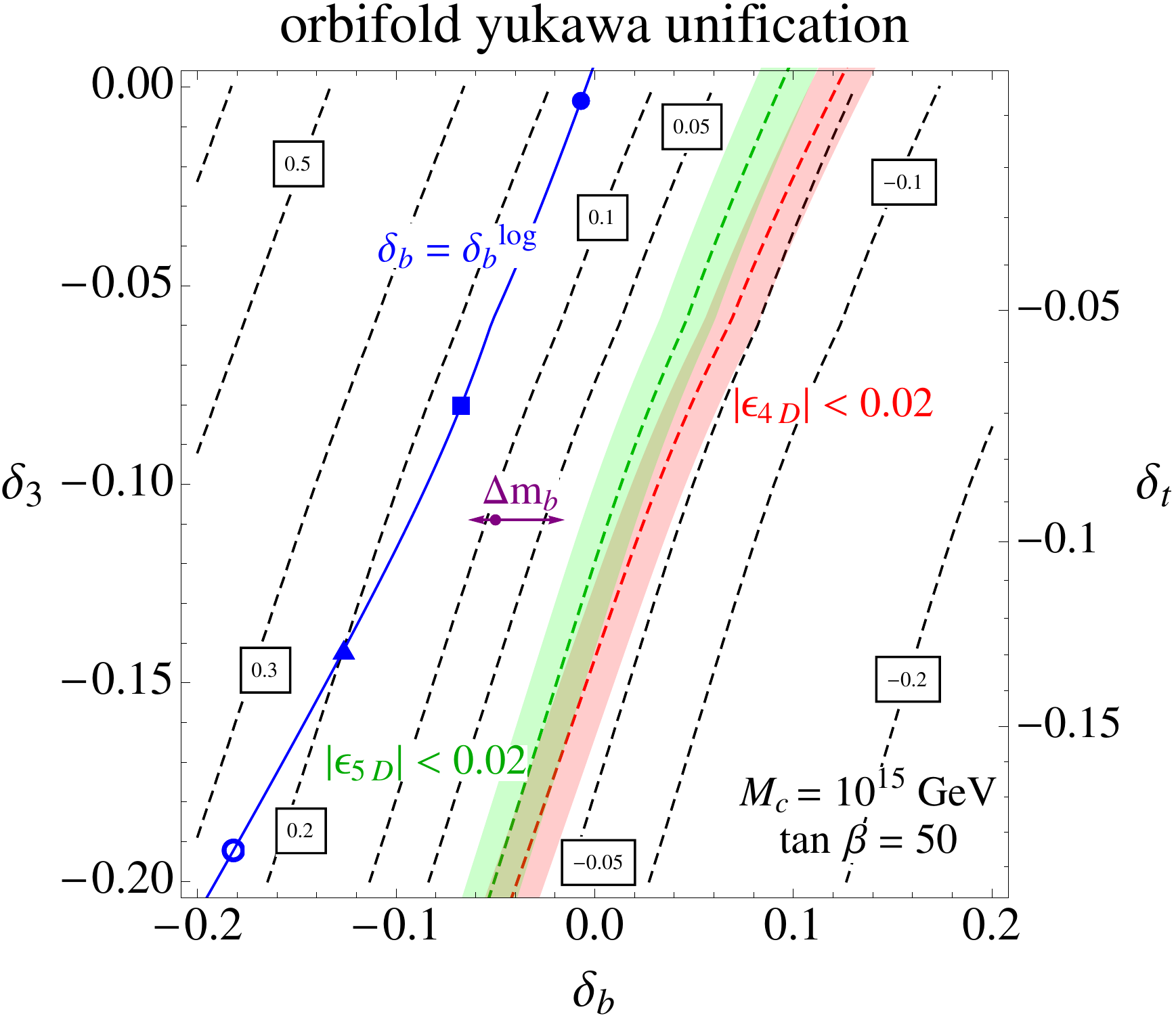} \end{center}
\caption{ \label{fig:orbi}
Yukawa unification in an orbifold GUT with compactification scale $M_c = 10^{15}$~GeV\@.  The black contours show the $b-\tau$ unification quality, $\epsilon$, evaluated at the compactification scale, as a function of the $y_b$ and $g_3$ threshold corrections.  At each point, $\tan \beta$ is chosen so that $y_t = y_b$ at the compactification scale, and for simplicity we set the $g_{1,2}$ and $y_\tau$ thresholds to 0.  The green shaded region denotes where $|\epsilon| < 0.02$ in the orbifold GUT\@.  For comparison, the red shaded region shows where $|\epsilon| < 0.02$ in the 4D GUT, as in figure~\ref{fig:ModelIndep_tbtau}.
}
\end{figure}

In the simplest 5d orbifold GUT \cite{Hall:2002ci} KK corrections improve gauge coupling unification compared to 4d unification, and reduce the unification scale to $10^{15}$ GeV\@.  The $b$ and $\tau$ couplings must therefore unify at this scale with $\epsilon = 0$, as shown for $\tan \beta$ near 50 by the green dashed line in figure~\ref{fig:orbi}.  Similar figures apply for lower $\tan \beta$.  Thus $\delta_b^{fin} \sim 0.1 - 0.15$ is an absolute necessity in such theories.  
 
\subsubsection{Summary}
\label{subsubsec:summary}

We have argued that there is a wide class of unified theories having small boundary condition and threshold corrections to the $b/\tau$ Yukawa ratio at the unified scale, $|\epsilon| < 0.02$.  Our analysis of supersymmetric threshold corrections in section~\ref{sec:susy_tbt} and the resulting collider, dark matter and B meson physics of section~\ref{sec:signals} applies to these theories.  Nevertheless, it is possible to construct theories where $\epsilon$ is much larger.  Such unified theories are required if supersymmetry is broken to the SM at the unified scale, when $\epsilon \sim 0.6$ is needed.

\section{MSSM Spectrum from Yukawa Unification and the Higgs mass}
\label{sec:susy_tbt}

In this section, we determine which supersymmetric spectra allow for third generation Yukawa unification and a Higgs mass of 125 GeV\@.  In order to remove dependency on unknown GUT physics, we simply assume that the GUT-scale threshold corrections to $y_b / y_\tau$ are small enough to be neglected, $|\epsilon| \lesssim 0.02$.  This assumption, which is motivated by the observation that the gauge thresholds are also this small, applies to a broad class of GUT models as described in section~\ref{sec:unifiedscale}.  Then, the success or failure of Yukawa unification depends entirely on the weak-scale SUSY spectrum, which must be chosen so that $\delta_b^{fin} \approx 0.1-0.2$, as we saw in section~\ref{sec:ModIndep}.  The program of this section is to determine which spectra satisfy this criterion.   We will also impose the requirement that the LSP abundance not overclose the universe, which when taken together with Yukawa unification implies, as we will see, an upper bound on the scalar masses.

The most important threshold corrections for $y_b / y_\tau$ are the corrections to $y_b, y_t$, and $g_3$, as we discussed in section~\ref{sec:ModIndep}.  For this section, we use the complete 1-loop SUSY threshold corrections, as computed by Ref.~\cite{Pierce:1996zz}.  The qualitative behavior of these thresholds is simple to understand.  The $g_3$ threshold is defined to be logarithmic only, in the $\overline{DR}$ scheme, and the $y_t$ threshold is dominated by the $\log$ contribution.  Therefore, both $\delta_t$ and $\delta_3$ are negative, with size determined logarithmically by the overall SUSY mass scale.   The $y_b$ threshold is a sum of the negative logarithmic piece and a $\tan \beta$ enhanced finite piece, $\delta_b^{fin}$.  As we found in section~\ref{sec:ModIndep}, the $b$ and $\tau$ Yukawas unify when $\delta_b^{fin} \approx 0.1-0.2$, where $\delta_b^{fin}$ is dominated by the gluino and chargino diagrams shown in figure~\ref{fig:bthresh_feyn}, which is approximately given by equation~\ref{eq:bthresh}.  The most important SUSY parameters for determining $\delta_b^{fin}$, and therefore whether or not the Yukawas unify, are,

\be
m_{Q_3}, \, m_{U_3}, \, m_{D_3}, \, A_t, \, M_3, \, \mu, \, \tan \beta.
\ee
Note that throughout this paper, we use the SUSY Les Houches Accord (SLHA) conventions for the signs of these parameters~\cite{Skands:2003cj, Allanach:2008qq}.

We choose to take seriously the hint, from ATLAS~\cite{ATLAS:2012ae} and CMS~\cite{Chatrchyan:2012tx}, that $m_h \approx$~125 GeV\@.  We will check to see that Yukawa unification can be made consistent with a Higgs boson of this mass.  Conveniently, the requirement that the correct Higgs mass is generated radiatively by the SUSY spectrum allows us to reduce the SUSY parameter space by one dimension.  Recall that the tree-level Higgs mass is given, in the MSSM, by $m_h^2 = \cos^2 2 \beta \, \, m_Z^2 \approx m_Z^2$ (where we have taken the decoupling limit, $m_A \gg m_Z$, and used the fact that in the parameter space of interest for Yukawa unification, $\tan \beta$ is large enough that $\cos^2 2 \beta \approx 1$).  Radiative corrections are required to raise the Higgs mass from $m_Z$ to 125~GeV\@.  The leading 1-loop correction is given by,
\be \label{eq:higgs}
m_h^2 \approx \cos^2 2 \beta \, \, m_Z^2 + \frac{3}{4 \pi^2} \, y_t^4 \, v^2 \left[ \log \frac{m_{\tilde t}^2}{m_t^2} + \frac{X_t^2}{m_{\tilde t}^2}\left( 1 - \frac{X_t^2}{12 m_{\tilde t}^2} \right)  \right] \approx 125~\mathrm{GeV},
\ee
where $v \approx 174$~GeV, $m_{\tilde t}^2 = m_{Q_3} m_{U_3}$, and $X_t = A_t - \mu/ \tan \beta$ determines the amount of left-right stop mixing.  By solving equation~\ref{eq:higgs}, we can determine $A_t$ for a given value of $m_{\tilde t}$.

\begin{figure}[t]
\begin{center} \includegraphics[width=0.5\textwidth]{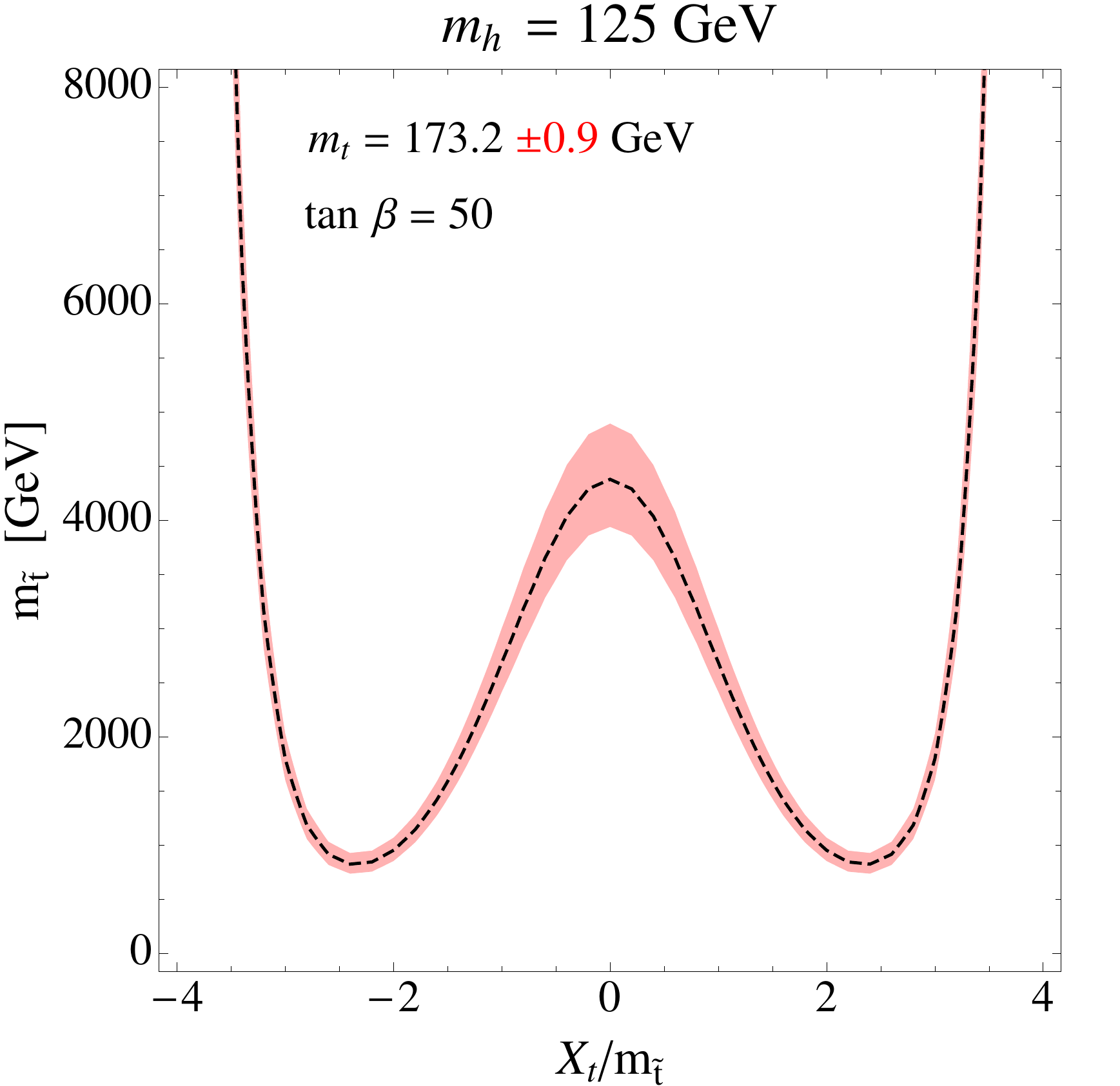}  \end{center}
\caption{\label{fig:mh125}
The values of the stop mass, $m_{\tilde t}^2 = m_{Q_3} m_{U_3}$ and stop mixing, $X_t$, that lead to $m_h = 125$~GeV, with $\tan \beta = 50$.  The red band shows the uncertainty coming from the top mass, $m_t = 173.2 \pm 0.9$~GeV\@.  Here the Higgs mass is calculated at 1-loop, using a numerical code~\cite{Bernal:2007uv} that allows for heavy stops by re-summing  $\log (m_{\tilde t} / m_Z)$.
}
\end{figure}

In practice, the RG improvement of equation~\ref{eq:higgs} must be taken into account to achieve accurate results.  In order to determine the Higgs mass as a function of the soft parameters, we use a version of Suspect designed to correctly handle heavy scalars, provided to us by the authors of Ref.~\cite{Bernal:2007uv}.  In this code, the Higgs quartic coupling $\lambda$ is run from the weak scale to the stop mass, $m_{\tilde t}$, using 1-loop RG and 1-loop thresholds computed both at the weak scale and the stop mass scale.  The result for $\tan \beta = 50$ is shown in figure~\ref{fig:mh125}, as a function of $m_{\tilde t}$ and $X_t / m_{\tilde t}$.  The red band shows the uncertainty coming from  the top mass, $m_t = 173.2 \pm 0.9$~GeV~\cite{Lancaster:2011wr}.  For small values of the $A$-term, $X_t\approx0$, we require $m_{\tilde t} \approx 4-5$~TeV\@.  The smallest value of the stop mass corresponds to the regime known as {\it maximal mixing}, where $X_t$ is chosen to maximize the second term in the brackets of equation~\ref{eq:higgs}, $X_t / m_{\tilde t} = \pm \sqrt{6}$.  Here, we find $m_{\tilde t} \approx 900$~GeV.\footnote{Note that near maximal mixing, the 2-loop contributions to $m_h$ are important for determining the stop mass and there remains a large amount of uncertainty on the necessary stop mass, with different estimates ranging from $m_{\tilde t} \sim 700-1000$~GeV, see Ref.~\cite{Hall:2011aa}.  We do not include the 2-loop effects in this paper, since the public codes implementing the 2-loop Higgs mass~\cite{Djouadi:2002ze, Frank:2006yh}  are unreliable in the limit of heavy scalars.}  Moving to larger values of $|X_t|$, the scalars drift up to heavier masses.  For $X_t / m_{\tilde t} = \pm \sqrt{12}$, the $X_t$ dependent terms of equation~\ref{eq:higgs} cancel against each other.  For larger values of $|X_t|$, the $X_t$ dependent terms are negative, and can be chosen to cancel against the $\log$, allowing for heavier scalars.

We have seen that $m_h \approx 125$~GeV potentially allows for very heavy scalars for special, large, values of $A_t$.   In this part of parameter space, the large $A_t$ leads to the presence of a lower energy charge and color breaking vacuum~\cite{Claudson:1983et}, and our vacuum is metastable.  The criterion for global stability is $A_t ^2+ 3 \mu^2 < 3 (m_{Q_3}^2 + m_{U_3}^2)$~\cite{Casas:1995pd}.  If we take $m_{Q_3} = m_{U_3} = m_{\tilde t}$ and assume that $\mu \ll A_t, m_{\tilde t}$, then metastability corresponds precisely to $X_t$ past maximal mixing, $|X_t| \gtrsim \sqrt{6} \, m_{\tilde t}$, where we have used that $X_t \approx A_t$ at large $\tan \beta$.  Splitting $m_{Q_3}$ from $m_{U_3}$, while keeping the Higgs mass fixed, has the effect of increasing the globally stable region.  For $X_t$ past maximal mixing, one must check whether or not the lifetime of our vacuum is long enough to support the present Age of the Universe.  The numerical analysis of Ref.~\cite{Kusenko:1996jn} found that our vacuum is sufficiently long-lived for $A_t ^2+ 3 \mu^2 \lesssim 7.5 (m_{Q_3}^2 + m_{U_3}^2)$.  For degenerate $m_{Q_3}$ and $m_{U_3}$, this would correspond to $X_t / m_{\tilde t} \lesssim \sqrt{15}$, or $m_{\tilde t} \lesssim 54$~TeV for $m_h = 125$~GeV\@.  Unfortunately, the numerical analysis of Ref.~\cite{Kusenko:1996jn} was only performed for $m_{\tilde t} \lesssim 2$~TeV and needs to be extended to include the part of parameter space relevant for $m_h \approx 125$~GeV\@.  We leave this for future work, and for now we tentatively consider the entire parameter space that gives $m_h = 125$~GeV\@.

\begin{figure}[t]
\hbox{ \hspace{-2cm}\includegraphics[height=10cm]{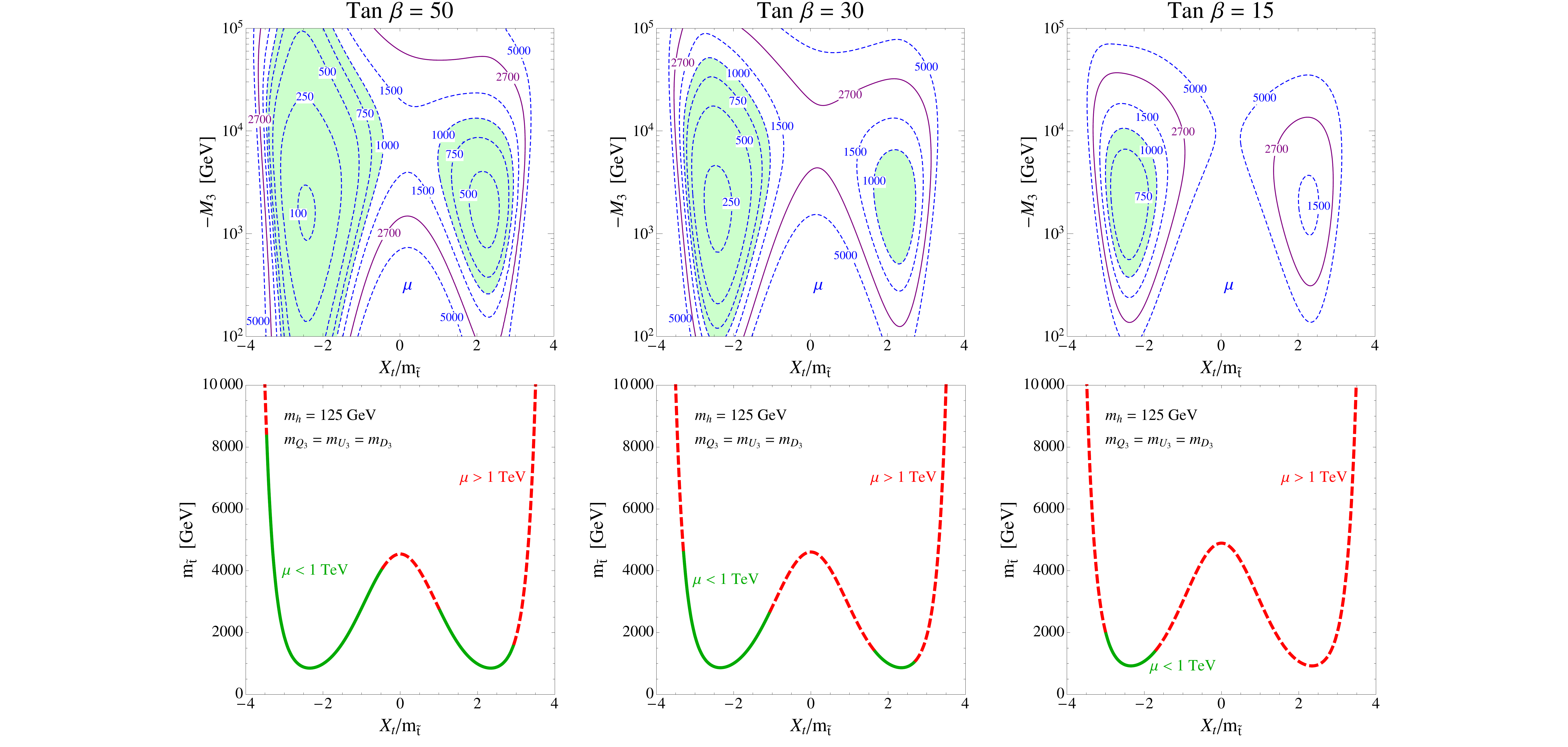} }
\caption{\label{fig:master_tanb}
The SUSY parameter space with third generation Yukawa unification, assuming degenerate stops and sbottoms $m_{Q_3} = m_{U_3} = m_{D_3}$, for $\tan \beta = 50, 30, 15$.  If $\tan \beta = 50$, then $t-b-\tau$ unification is possible; for the lower values of $\tan \beta$ only $b-\tau$ unification is possible.  We determine the stop mass as a function of the stop mixing, $X_t$, by demanding $m_h = 125$~GeV, as shown in the lower row of plots.  In the upper plots, we show the value of $\mu$ that gives Yukawa unification as a function of $X_t$ and $M_3$.  The green shaded area has $\mu < 1$~TeV, allowing for a bino-Higgsino LSP whose energy density does not overclose the universe.  If the LSP is wino-like and constitutes all of DM, then $M_2 \approx 2.7$~TeV and we require $\mu > M_2$, which is the area outside the purple contour.
}
\end{figure}

We show our primary result in figure~\ref{fig:master_tanb}: the SUSY parameter space that allows for precise $b/\tau$ unification.  As discussed above, we have related $m_{\tilde t}$ and $A_t$ by imposing $m_h \approx 125$~GeV\@.  For simplicity, we take degenerate stops and sbottoms (we will relax this assumption below), $m_{\tilde t} = m_{Q_3} = m_{U_3} = m_{D_3} $.  The top row of plots show the value of $\mu$ that leads to $b-\tau$ unification, as a function of $X_t / m_{\tilde t}$ and $M_3$, for $\tan \beta = 50, 30, 15$.  The first value, $\tan \beta = 50$, leads to full $t-b-\tau$ unification, as we saw in figure~\ref{fig:ModelIndep_tbtau}, while the lower values of $\tan \beta$ allow for $b-\tau$ unification only.  The bottom row of plots shows the corresponding value of $m_{\tilde t}$, as a function of the $X_t / m_{\tilde t}$ axis, that gives $m_h = 125$~GeV\@.  The remaining SUSY parameters have sub-leading effects on Yukawa unification, but to be definite, we fix $M_{1,2}$ using the gaugino unification relation, $6 M_1 \approx 3 M_2 \approx M_3$, all of the other sfermions are fixed to $m_{\tilde t}$, and we take $m_A = 3$~TeV\@.

The interplay between Yukawa unification and the cosmological LSP abundance is key.  As stated in the introduction, we consider high-scale mediation and conserved $R$ parity so that the LSP is stable and has an abundance in standard cosmology given by thermal freezeout.  The shaded green region with $\mu <$ TeV is the one of primary concern to us, since it corresponds to the simplest DM possibility of bino/Higgsino LSP\@.  Note that a pure Higgsino has the correct abundance, $\Omega_{cdm} h^2 = 0.11$~\cite{pdg}, when $\mu \approx 1$~TeV\@.  A bino-Higgsino admixture does not overclose, $\Omega_{\tilde N_1} < \Omega_{cdm}$, as long as $\mu <  1$~TeV and $\mu \lesssim M_1$, and a bino-Higgsino admixture typically has the correct abundance to constitute all of dark matter in the so-called well-tempered regime, $\mu \sim M_1$~\cite{ArkaniHamed:2006mb}.  Since the precise value of $M_1$ has little effect on Yukawa unification, anywhere within the green region can be made to avoid overclosure simply by choosing $M_1 \gtrsim \mu$.

There are several other possibilities for SUSY dark matter.  A bino/Higgsino admixture can have $\mu > 1$~TeV and still avoid overclosure if DM co-annihilates with a slepton or squark or if DM annihilates through a heavy Higgs pole.
Both of these options necessitate extra light scalars and require a tuning among masses to avoid overclosure.  We will not consider co-annihilation or heavy Higgs pole further in this paper.  Another option, that does not necessitate tuning parameters to avoid overclosure, is that the LSP is wino-like, with $M_2 < M_1, \mu$.  Then the correct abundance for the wino to be all of DM is achieved for $M_2 \approx 2.7$~TeV~\cite{Hisano:2006nn, Cirelli:2007xd}, and lighter winos would contribute a subdominant portion of the energy density.  If we assume that a wino LSP contributes all of DM and demand Yukawa unification then we must have $\mu > 2.7$~TeV: this boundary is shown as a purple line in figure~\ref{fig:master_tanb}.

Returning to the simplest possibility: that DM is a bino/Higgsino admixture, we see from figure~\ref{fig:master_tanb}, that avoiding overclosure, $\mu < 1$~TeV, leads to an upper bound on the stop mass when we demand Yukawa unification.   For $\tan \beta = 50$, as is appropriate for $t-b-\tau$ unification, we find that $m_{\tilde t} \lesssim 2.8$~TeV is required when $A_t$ and $M_3$ have opposite signs, which corresponds to the right half of the plane shown in the figure.  Here, the gluino and chargino diagrams have opposite signs: see equation~\ref{eq:bthresh}.  We note that a relative sign between $A_t$ and $M_3$ is favored by the renormalization group because $M_3$ contributes radiatively to $A_t$, driving it to have an opposite sign, 
\be \label{eq:AtRG}
\frac{d A_t}{dt} = \frac{1}{8 \pi^2}\left(\frac{16}{3} g_3^2 M_3 + 6 |y_t|^2 A_t + |y_b|^2 A_b + \ldots \right).
\ee
In order for $A_t$ and $M_3$ to have the same sign, there must be a large $A$-term in the UV to compensate the radiative contribution from $M_3$.  In this case, the two diagrams add  and heavier stops are possible, $m_{\tilde t} \lesssim 8500$~GeV\@.  Since the thresholds are proportional to $\tan \beta$, lowering $\tan \beta$ has the effect of shrinking the allowed parameter space and necessitating even lighter stops (also, recall from figure~\ref{fig:ModelIndep_btau} that a larger $\delta_b^{fin}$ is required at lower $\tan \beta$).  For $\tan \beta = 30$, we require $m_{\tilde t} \lesssim 1400$~GeV when $A_t$ and $M_3$ have opposite signs, and $m_{\tilde t} \lesssim 4600$~GeV when they have the same sign.  For $\tan \beta = 15$,  the region where $A_t$ and $M_3$ have opposite signs is removed completely, and we require $m_{\tilde t} \lesssim 2000$~GeV when the signs are the same.  We find no $b-\tau$ unification at all when $\tan \beta \lesssim 10$.  We note that with our sign conventions, nearly the entire parameter space with Yukawa unification and $\mu < 1$~TeV is where $M_3$ is negative, so we have only shown this sign choice in figure~\ref{fig:master_tanb}.

\begin{figure}[t]
\hbox{ \hspace{0.5cm}\includegraphics[height=10cm]{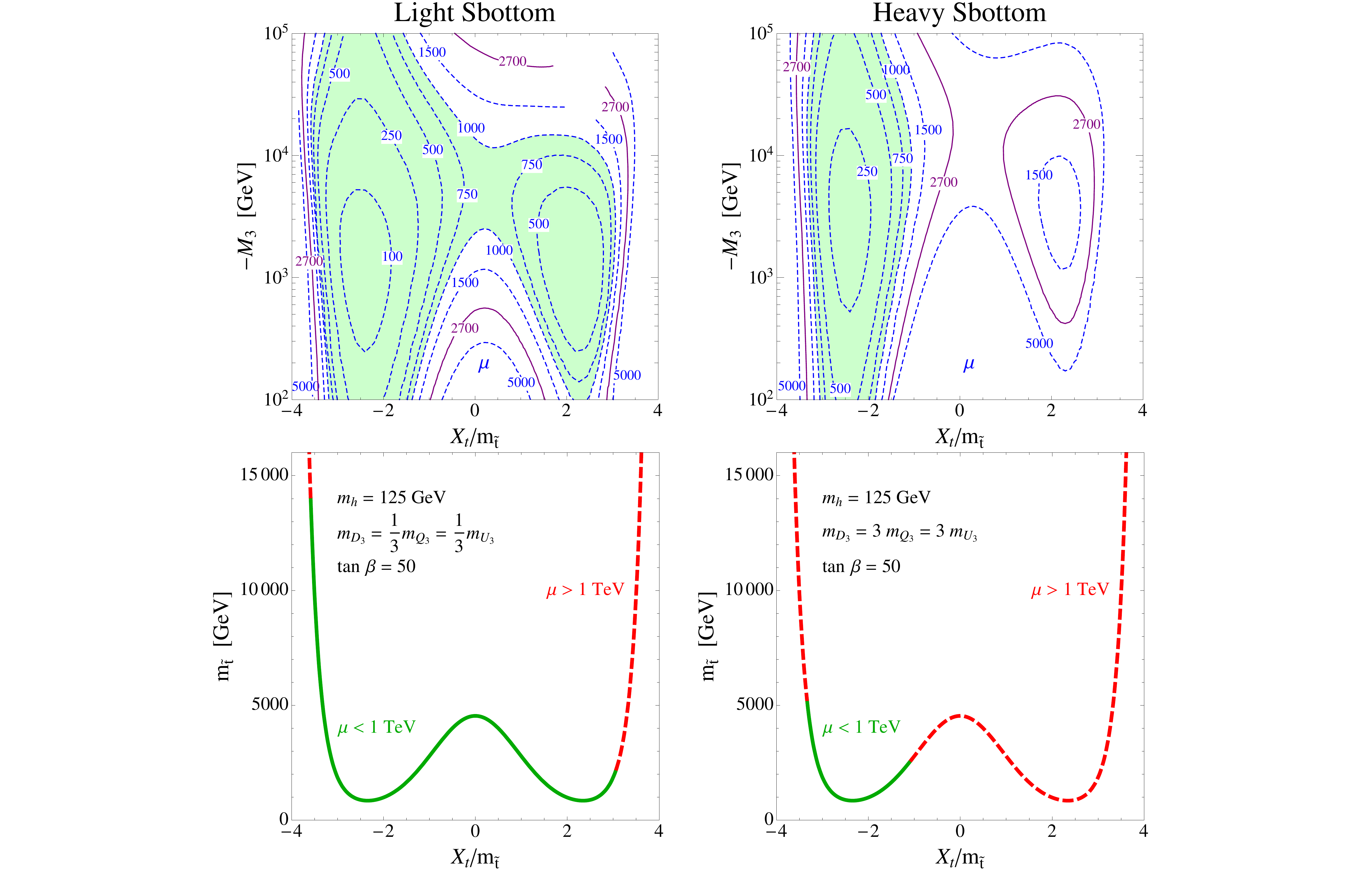} }
\caption{ \label{fig:master_sbottom}
The allowed parameter space for Yukawa unification, as in figure~\ref{fig:master_tanb}, except relaxing the assumption of degenerate stops and sbottoms.  On the left, we consider lighter right-handed sbottoms, with $m_{D_3} = 1/3 \, m_{\tilde t}$ and $m_{\tilde t} = m_{Q_3} = m_{U_3}$.  On the right, we consider heavier right-handed sbottoms, $m_{D_3} = 3 \, m_{\tilde t}$.  Lighter sbottoms enhance the size of the gluino diagram of figure~\ref{fig:bthresh_feyn}, leading to a larger region with Yukawa unification, while heavier sbottoms deplete the gluino diagram and shrink the allowed region.
}
\end{figure}

So far, we have assumed that the stops and sbottoms are degenerate.  We relax this assumption in figure~\ref{fig:master_sbottom}.  On the left, we show a case with lighter right-handed sbottom, $m_{D_3} = 1/3 \, m_{\tilde t}$, where $m_{\tilde t} = m_{Q_3} = m_{U_3}$.  On the right, we show the case with heavier right-handed sbottom, $m_{D_3} = 3 \, m_{\tilde t}$.  The effect of splitting the stops from the sbottoms is to change the relative size of the gluino diagram, which has sbottoms propagating in the loop, to the chargino diagram, which has stops propagating in the loop.  When the sbottoms are lighter, as in the left diagram, the size of the gluino diagram is enhanced, leading to a larger region with Yukawa unification.  In this case, stops as heavy as $14$~TeV allow for Yukawa unification when $A_t$ and $M_3$ have the same sign.  However, this implies that the sbottom is lighter than 5 TeV, such that the requirement of one (somewhat) light scalar has not been circumvented.  When the sbottoms are heavier, as in the right diagram, the gluino diagram is depleted.  In this case, we find no Yukawa unification when $A_t$ and $M_3$ have opposite sign, and when they have the same sign we require $m_{\tilde t} \lesssim 5200$~GeV\@.

\section{Flavor and Dark Matter Phenomenology}
\label{sec:signals}

In this section we discuss possible experimental signatures of supersymmetry with Yukawa unification.  We found in section~\ref{sec:susy_tbt} that most of the parameter space of Yukawa unification (plus dark matter) has stops and sbottoms lighter than several TeV and therefore potentially accessible to experiments.  We discuss the flavor signals and constraints in section~\ref{subsec:Bmeson}. As we will see, the large values of $\tan \beta$  necessary for Yukawa unification lead to observable signals in $B$-meson decays.  In section~\ref{subsec:dm}, we discuss the phenomenology of dark matter within our parameter space.

Before we move on to discus flavor and dark matter, we briefly consider the exciting prospect of discovering superpartners at the LHC\@.  Yukawa unification implies that the stop and sbottom must be lighter than several TeV and therefore may be produced at the LHC\@.  The observability  is highly dependent on the precise spectrum.  The masses of the superpartners will determine the production cross-section, and their spectrum will determine which decays dominate and therefore whether or not low-background final states (preferably with leptons and $b$-jets) are populated.

\begin{figure}[t]
\begin{center} \includegraphics[width=0.5\textwidth]{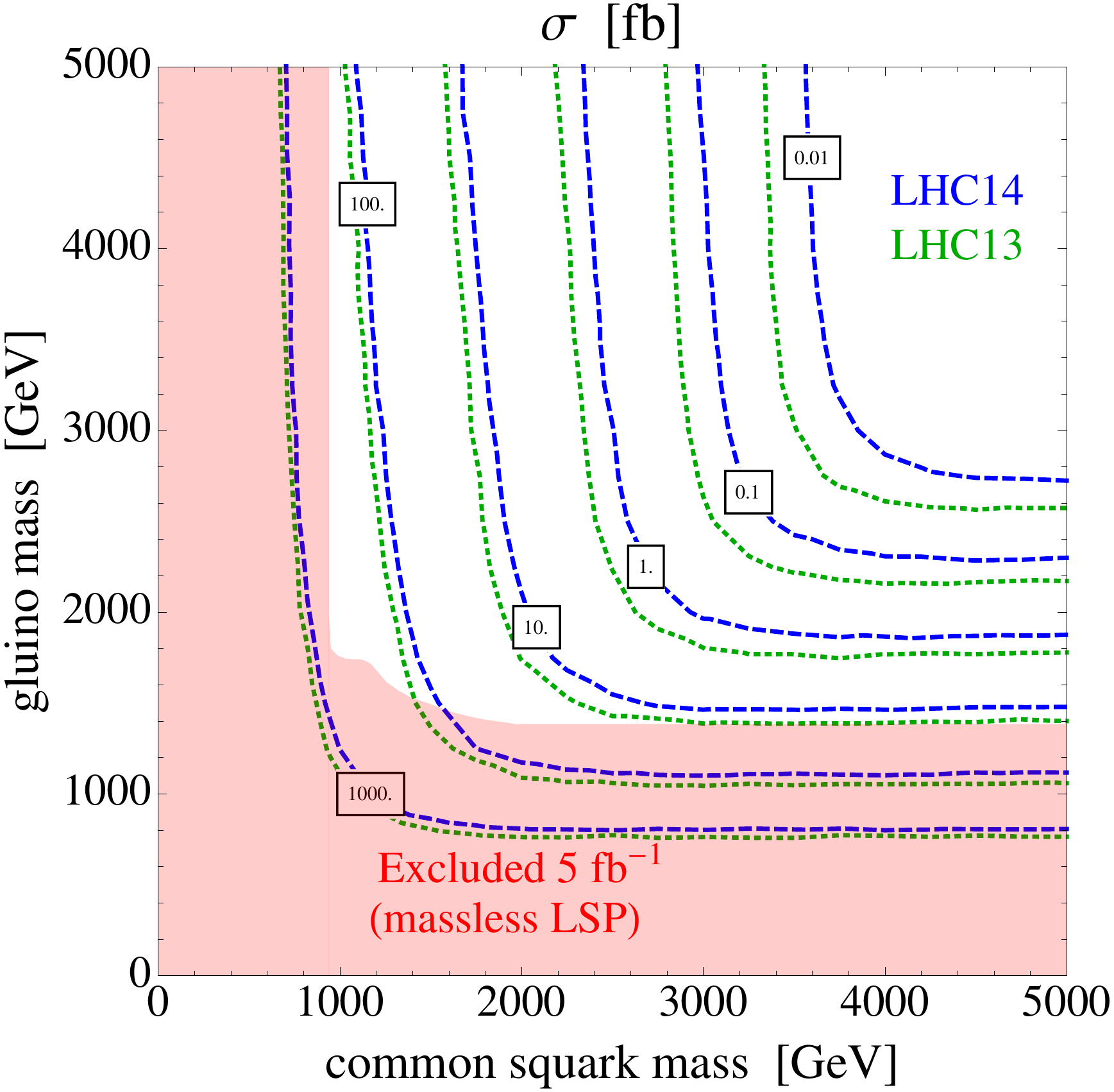}  \end{center}
\caption{ \label{fig:cross}
The total colored sparticle production cross-section, at LO, as a function of the gluino mass and a common squark mass~\cite{Sjostrand:2006za}.  The red area is already excluded by ATLAS, assuming the squarks and gluinos decay directly to a massless neutralino~\cite{ATLAS-CONF-2012-033}.
}
\end{figure}

As a simple example of the overall SUSY production rate, figure~\ref{fig:cross} shows the cross-section as a function of the gluino mass and a common squark mass, at center of mass energies of 13 and 14~TeV\@.  If a search in a low-background channel has an efficiency of $\sim 10\%$, then 50 events before cuts are more than enough for a discovery.  For degenerate gluino and squarks, this corresponds to a 100~fb$^{-1}$ reach of 2.4 (2.3) TeV at 14 (13) TeV\@.  It is important to keep in mind that discovery is not guaranteed because while Yukawa unification requires a light stop and sbottom, it does not require a light gluino (see section~\ref{sec:susy_tbt}) and the masses of the first and second generation squarks are unconstrained.  Still, assuming a simple scenario for supersymmetry breaking, the stop and sbottom masses are probably near the masses of the other colored sparticles.  Therefore, we find it highly encouraging that the eventual LHC energy will be enough to probe most superpartner scales relevant for Yukawa unification.

\subsection{$B$-meson decays}
\label{subsec:Bmeson}

In this section we consider the flavor violating signals of SUSY with Yukawa unification.  We found in section~\ref{sec:susy_tbt} that the favored parameter space of Yukawa unification has stops and sbottoms with masses in the $\sim 1-10$~TeV range.  For superpartners at these scales, generic squark soft masses are forbidden by $\Delta F = 2$ processes such as $K- \bar K$ mixing.  Therefore the soft masses must possess a special flavor structure, such as Minimal Flavor Violation (MFV)~\cite{Chivukula:1987py, Hall:1990ac, D'Ambrosio:2002ex} or a $U(2)$ flavor symmetry~\cite{Pomarol:1995xc, Barbieri:1995uv, Barbieri:1997tu} (or $U(2)^3$~\cite{Barbieri:2011ci}).  Even if we do assume MFV, there are $B$-meson decays that receive $\tan \beta$ enhanced contributions~\cite{Altmannshofer:2010zt}.  This is highly relevant for Yukawa unification, because we found above that $t-b-\tau$ unification requires $\tan \beta \approx 50$ and $b-\tau$ unification plus dark matter requires $\tan \beta > 10$.  For now, we assume that CP phases are small and only consider flavor violating effects.  We will comment on the limits on CP violation at the end of this section.

\begin{figure}[t]
\begin{center} \includegraphics[width=0.7\textwidth]{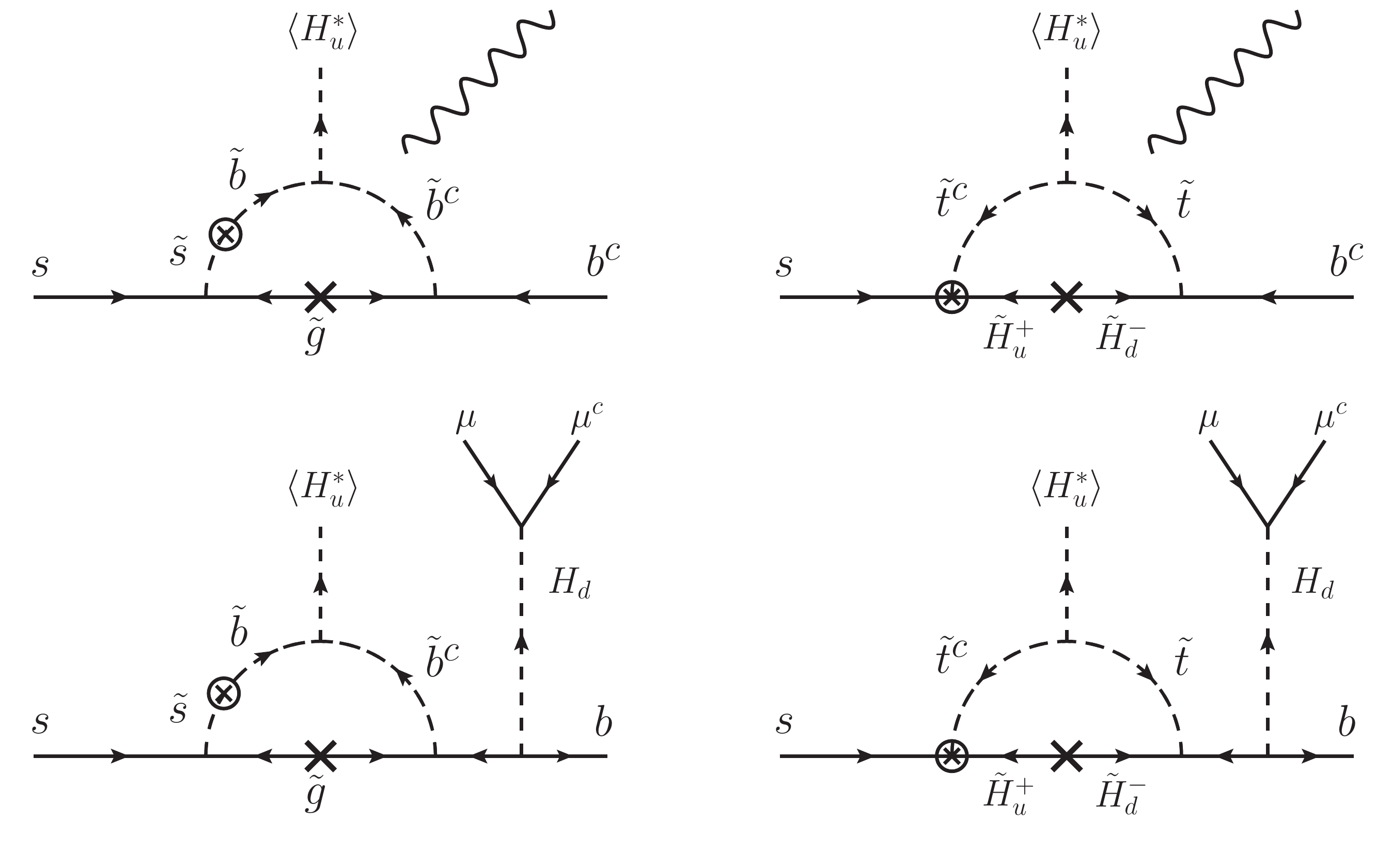}  \end{center}
\caption{\label{fig:flavor_feyn}
The leading, $\tan \beta$ enhanced, diagrams contributing to  $b \rightarrow s \gamma$ and $B_s \rightarrow \mu^+ \mu^-$ are shown above and below, respectively.  There is a close relationship between the size of these diagrams and the finite threshold correction to the bottom mass (see figure~\ref{fig:bthresh_feyn}).  In the SUSY-CKM basis, the leading contributions of the gluino exchange diagrams on the left are proportional to left-left down squark mass mixing, ${(\delta_d^{LL})}_{32}$, while the Higgsino exchange diagrams on the right are proportional to $V_{ts}$, coming from the Higgsino vertex.
}
\end{figure}

The most constraining processes are $b \rightarrow s \gamma$ and $B_s \rightarrow \mu^+ \mu^-$.  The leading SUSY diagrams, at large $\tan \beta$, arise from gluino and Higgsino exchange, and are shown in figure~\ref{fig:flavor_feyn}.  There is a tight relationship between the amplitudes for these processes and the threshold correction to the bottom mass, which is generated by similar 1-loop diagrams, as can be seen by comparing figures~\ref{fig:bthresh_feyn} and~\ref{fig:flavor_feyn}.  Limits from $B$-meson decays are discussed in many papers on Yukawa unification, see for example Refs.~\cite{Rattazzi:1995gk, Blazek:2001sb, Tobe:2003bc, Altmannshofer:2008vr}.  These references all consider unified scalar masses at the GUT scale.  Our approach differs because we allow for general soft terms at the weak scale, without specifying the high-scale boundary condition.

We now consider the flavor structure of these diagrams.  We work in the basis where the fermion masses and squark soft masses have been diagonalized, with conventions described in appendix~\ref{app:flavor}\@.
In the limit of small left-right mixing, this corresponds to the squark mass eigenstate basis, and all flavor violation can be written in terms of the Cabibbo-Kobayashi-Maskawa (CKM) matrix and the gluino vertices, $(W^q)^{ij} \, \tilde q_i^\dagger q_j \tilde g$ (see Ref.~\cite{Nir:1993mx} and appendix~\ref{app:flavor}).  Small left-right mixing is a good approximation in most of our parameter space, because the squarks are heavier than 1 TeV and left-right mixing is suppressed by $v / \tilde m$.

First consider the Higgsino diagrams contributing to $b \rightarrow s \gamma$ and $B_s \rightarrow \mu^+ \mu^-$.  The diagrams for the two processes are proportional to the same flavor factor, but involve different loop functions.   Assuming that the Higgsino diagrams are dominated by stop exchange, because of the $A$-term insertion,
\be
i M_{\tilde H} \propto y_b \, y_t \, A_t V_{ts} (W^{u^c}_{33} W^d_{33})^* \approx y_b \,y_t \, A_t V_{ts},
\ee
where the second step assumes small mixing, so that the 3-3 entries of $W^{u^c}$ and $W^d$ are close to 1.  We see that the Higgsino diagrams have an irreducible contribution proportional to $V_{ts}$, arising from the Higgsino vertex connecting a strange quark and top squark.

Next we turn to the gluino diagrams.  The leading contribution has the flavor structure,
\be \label{eq:gluinoflavor}
i M_{\tilde g} \propto \left( {W^d}^T_{2j} \, P_j \, {W^d}^*_{j3} \right) y_b \left( {W^{d^c}}^\dagger_{33} \,  P^c_{3} \, W^{d^c}_{33}  \right) \approx y_b P^c_{3} \left(W^d_{32} P_3 + {W^d}^*_{23} P_2 \right)
\ee
where $P_i$ ($P_3^c$) denotes the propagator of $\tilde d_i$ ($\tilde d^c_3$) and the second step assumes 1-3 mixing is small enough to be neglected.  Note that this diagram vanishes for degenerate squarks because of a super-GIM mechanism that follows from the unitarity of $W^d$.
For non-degenerate squarks, the diagram is proportional to $W^d_{23}$, which is an unknown factor of order $V_{ts}$ in theories with MFV\@.  

It is also common to describe the flavor violation in the so-called SUSY-CKM basis, where the entire quark superfields are rotated into the basis where the fermion masses are diagonal.  In this basis, it is useful to work in the mass-insertion-approximation when flavor violation is small, in which case the leading contribution from the gluino diagram is proportional to the left-handed 2-3 mixing, ${(\delta_d^{LL})}_{32}$.  The factor ${(\delta_d^{LL})}_{32}$ is related to  $W^d_{23}$ by equation~\ref{eq:RelateBases}, and is expected to be of order  $V_{ts}$ in theories with MFV\@.  In general, the gluino diagrams also contain terms proportional to the right-handed squark mixing, ${(\delta_d^{RR})}_{32}$.  This mixing is suppressed in theories with MFV and here we assume that it is subdominant to the contribution from left-handed squark mixing.

We now describe the experimental constraints, and reach, in some detail.  We will first discuss the constraint coming from $b \rightarrow s \gamma$, which has already been measured and agrees with the SM prediction.  Then we will discuss $B_s \rightarrow \mu^+ \mu^-$, which has not yet been measured and, as we will see, may have a rate that differs significantly from the SM prediction.

\begin{figure}[t]
\begin{center} \includegraphics[width=1\textwidth]{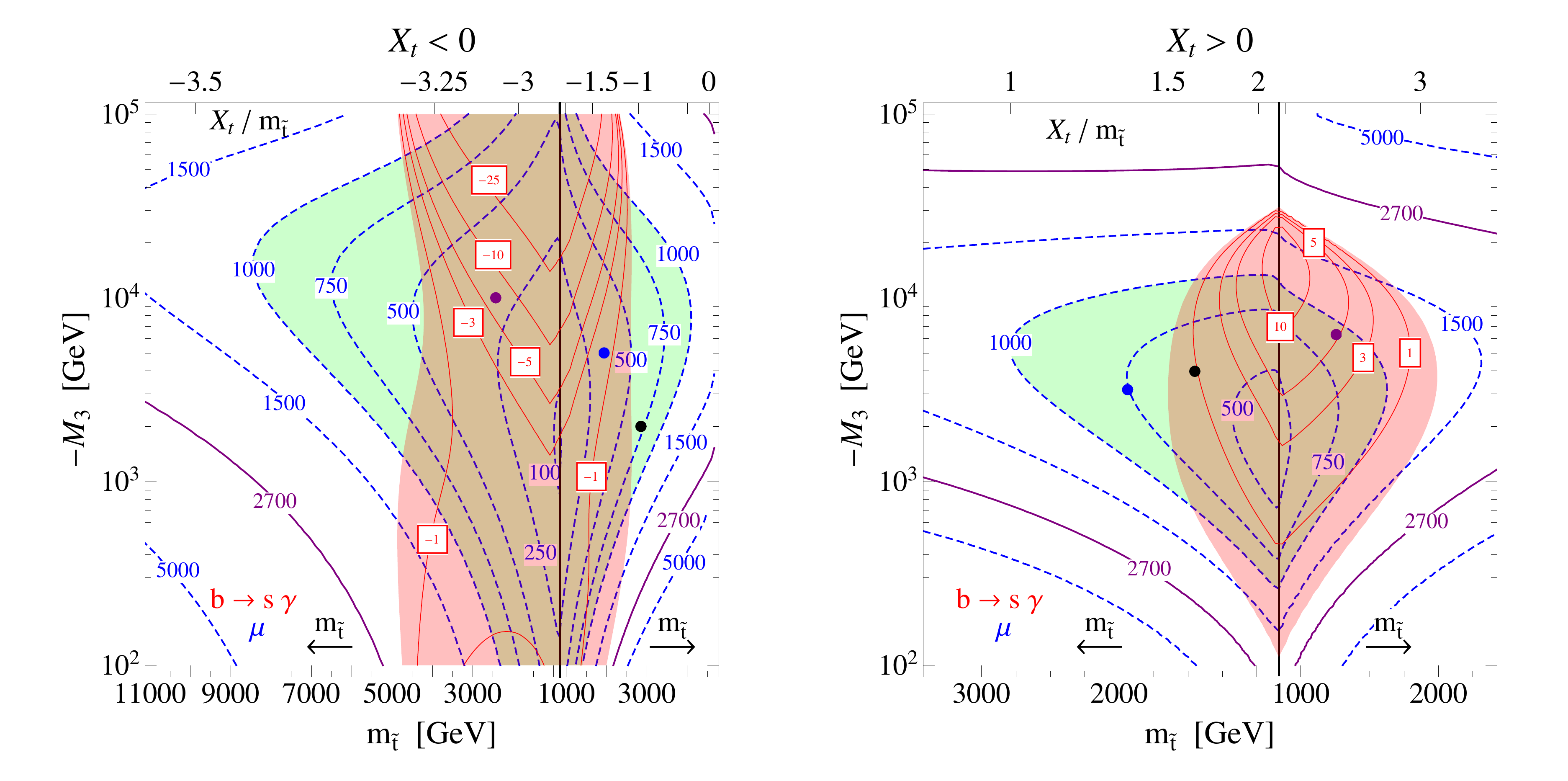}  \end{center}
\caption{ \label{fig:bsg2D}
The allowed region for Yukawa unification with $\tan \beta = 50$, with the limit from $b \rightarrow s \gamma$ overlaid.  The y-axis is the gluino mass and the x-axis corresponds both to the stop mixing, $X_t$ and the stop mass, which is non-monotonic in $X_t$ because after fixing $m_h = 125$~GeV, the stop mass increases moving away from maximal mixing, $X_t = \pm \sqrt 6 \, m_{\tilde t}$ (see figure~\ref{fig:mh125}).  As in figure~\ref{fig:master_tanb}, the blue contours indicate the value of $\mu$ necessary for Yukawa unification and the shaded green region is consistent with bino/Higgsino dark matter,  $\mu < 1$~TeV\@.  The red region is excluded, at $2\sigma$, by $b \rightarrow s \gamma$ if 2-3 down squark mixing is ignored, ${(\delta_d^{LL})}_{32} = 0$, such that only Higgsino exchange contributes (we take $m_A = 10$~TeV so that charged Higgs exchange is decoupled).  The red contours show the minimum value of $|{(\delta_d^{LL})}_{32}|$, in units of $|V_{ts}|$, necessary to bring $b \rightarrow s \gamma$ in accord with the observed branching ratio.  The black, blue, and purple points are considered in more detail in figures~\ref{fig:bsg} and~\ref{fig:BsMuMureach}.
}
\end{figure}

The branching ratio for $b \rightarrow s \gamma$ has been precisely measured by the B-factories to be $(3.55 \pm 0.26) \times 10^{-4}$~\cite{Asner:2010qj}, which agrees with the NNLO SM prediction of $(3.15 \pm 0.23)\times10^{-4}$~\cite{Misiak:2006zs}.  This presents a serious constraint on theories with large $\tan \beta$, even for squarks in the several TeV range, because the amplitude of the SUSY contribution is proportional to $\tan \beta$.  Figure~\ref{fig:bsg2D} shows the allowed region for Yukawa unification for $\tan \beta = 50$, as in figure~\ref{fig:master_tanb} except now shown as a function of the stop mass and gluino mass.  As before, the blue contours indicate the values of $\mu$ necessary for Yukawa unification, and the shaded green region has $\mu < 1$~TeV and is therefore compatible with a bino/Higgsino LSP that does not overclose the Universe.  The red shaded area would be excluded at $2 \sigma$ by $b \rightarrow s \gamma$ if we were only to consider the irreducible contribution from Higgsino exchange, by setting ${(\delta_d^{LL})}_{32} = 0$, which turns off the gluino diagram.  Note that for our numerical results, we for simplicity use the mass-insertion-approximation results of Ref.~\cite{Altmannshofer:2009ne}.

\begin{figure}[t]
\begin{center} \includegraphics[width=1\textwidth]{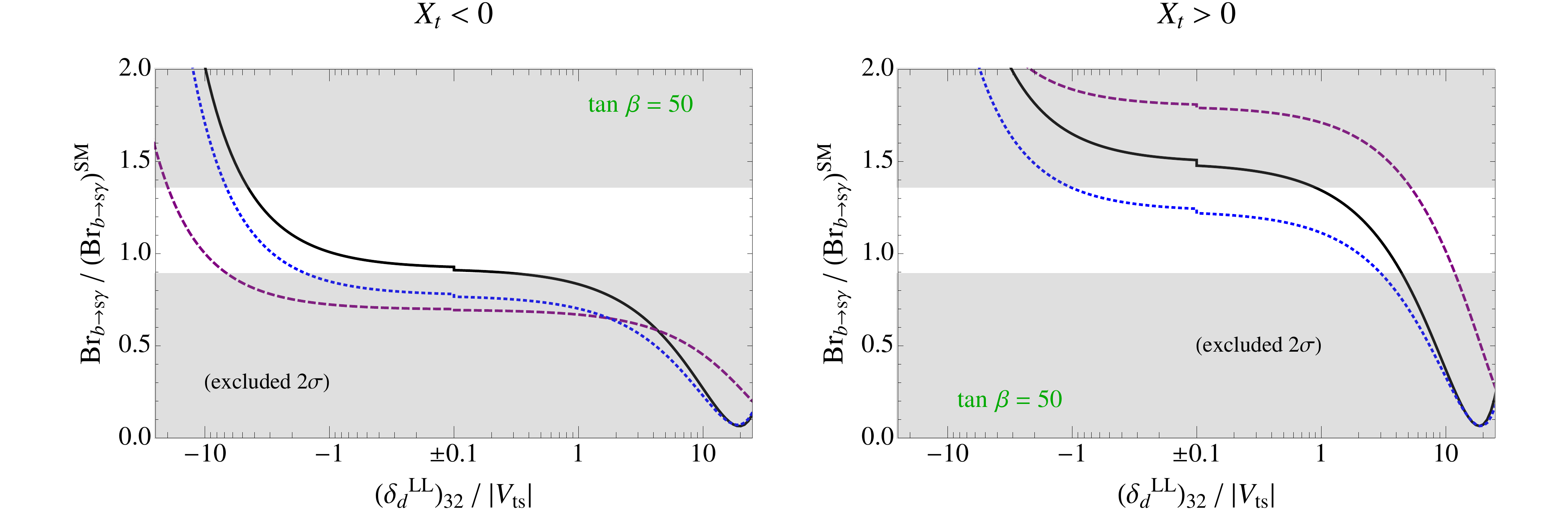}  \end{center}
\caption{\label{fig:bsg}
The branching ratio for $b \rightarrow s \gamma$ normalized to the SM prediction of $(3.15 \pm 0.23)\times10^{-4}$~\cite{Misiak:2006zs}, as a function of ${(\delta_d^{LL})}_{32} $ in units of $|V_{ts}|$.  The left (right) side of each plot corresponds to ${(\delta_d^{LL})}_{32} < 0$ ($>0$).  The black, blue, and purple curves correspond to the corresponding points in parameter space denoted on figure~\ref{fig:bsg2D}.  The shaded gray region is excluded by more than $2\sigma$.  Notice that for each point, the branching ratio of $b \rightarrow s \gamma$ is consistent with the limit for a wide range of values of ${(\delta_d^{LL})}_{32}$.  We have taken $m_A = 10$~TeV so that charged Higgs exchange is decoupled.
}
\end{figure}

If the Higgsino diagram were the end of the story, we would conclude that a large portion of the Yukawa-unified parameter space is already excluded.  However, it is important to include the gluino contribution, which depends on the unknown coefficient, ${(\delta_d^{LL})}_{32}$.  We show the branching ratio of $b \rightarrow s \gamma$ as a function of ${(\delta_d^{LL})}_{32}$ in figure~\ref{fig:bsg} for the points indicated by black, blue, and purple dots in figure~\ref{fig:bsg2D}.  The shaded area of the plot is excluded at $2\sigma$ and we see that each point is safe for a wide range of values of ${(\delta_d^{LL})}_{32}$.  We find that no fine-tuning is necessary because the Higgsino and gluino diagrams are of similar size and their sum is easily consistent with the constraint.  In figure~\ref{fig:bsg2D}, the red contours indicate the minimum absolute value of ${(\delta_d^{LL})}_{32}$ necessary for $b \rightarrow s \gamma$ to be consistent with observation, in units of $|V_{ts}|$.  In most of the parameter space, a value of $\mathcal{O}(V_{ts})$ is sufficient.

We make a few technical comments about $b \rightarrow s \gamma$ before moving on.  In addition to the gluino and Higgsino exchange diagrams, there is a also a diagram with wino exchange, where the wino mixes with a Higgsino.  This diagram is proportional to the same flavor structure appearing in the first factor of parentheses in equation~\ref{eq:gluinoflavor}.  The wino exchange is typically subdominant to gluino and Higgsino exchange, but we have included it in our numerical analysis by setting the ratio $M_3 : M_2 = 3$, as is approximately satisfied by spectra with gaugino unification.  Our answers are not very sensitive to the choice of $M_2$.  A second comment is that we have been focusing on the allowed region of $b \rightarrow s \gamma$ where the SUSY contribution is small.  There also used to be an allowed region where the SUSY contribution has the opposite sign of the SM contribution, but where their sum happens to have the same magnitude as the SM contribution.  This finely-tuned option is now excluded by LHCb measurements of $b \rightarrow X_s l^+ l^-$~\cite{Altmannshofer:2011gn}.  In the region we are interested in, the SUSY contribution to $b \rightarrow s \gamma$ is small, and in this limit $b \rightarrow X_s l^+ l^-$ is not relevant.

More interesting than $b \rightarrow s \gamma$, which is already observed to agree with SM, is the potential to see an observable deviation in the rate for $B_s \rightarrow \mu^+ \mu^-$.  This process is predicted to have the branching ratio $(3.2 \pm 0.2) \times 10^{-9}$ in the SM~\cite{Buras:2010mh}.  The experimental limits have been rapidly approaching this target, and LHCb now sets a 95\% limit of $4.5 \times 10^{-9}$~\cite{Aaij:2012ac}.  CMS is close behind with a 95\% limit of $7.7 \times 10^{-9}$~\cite{Chatrchyan:2012rg}.  It is likely that we will know very soon whether or not this rate agrees with the SM prediction or shows an observable enhancement or, in the case of destructive interference with SM diagrams, depletion.  The decay is generated by the 4-fermion operator,
\be
\mathcal{L}_{eff} \supset  C_s \, m_b \, \bar s_L b_R \bar \mu \mu, 
\ee
and also the operator with a $\gamma_5$ between the muons, and the corresponding operators derived by flipping parity, $L \leftrightarrow R$.

The leading SUSY contribution is mediated by a heavy Higgs and has an amplitude proportional to $\tan^3 \beta$~\cite{Babu:1999hn}.  This contribution follows directly from the flavor-violating analogue of the finite threshold correction to the bottom mass, which induces, in mass eigenstate, an $\bar s_L b_R H$ coupling.  Therefore, the SUSY contribution to the coefficient of this operator can be written directly in terms of the finite bottom threshold, dressed by the appropriate flavor factors,
\be \label{eq:BsMuMuAmp}
C_S^{SUSY} = \frac{m_\mu}{v^2} \frac{\tan^2 \beta}{m_A^2} \left[ {(\delta_d^{LL})}_{32} \, \delta_b^{\tilde g} + V_{ts}^* \, \delta_b^{\tilde H}  \right],
\ee
where $\delta_b^{\tilde g}$ and $\delta_b^{\tilde H}$ denote the gluino and Higgsino contributions to the finite bottom threshold, which are given by the two terms in equation~\ref{eq:bthresh}.  The size of the SUSY contribution is highly sensitive to $\tan \beta$ and to the value of the heavy Higgs mass.

\begin{figure}[t]
\begin{center} \includegraphics[width=1\textwidth]{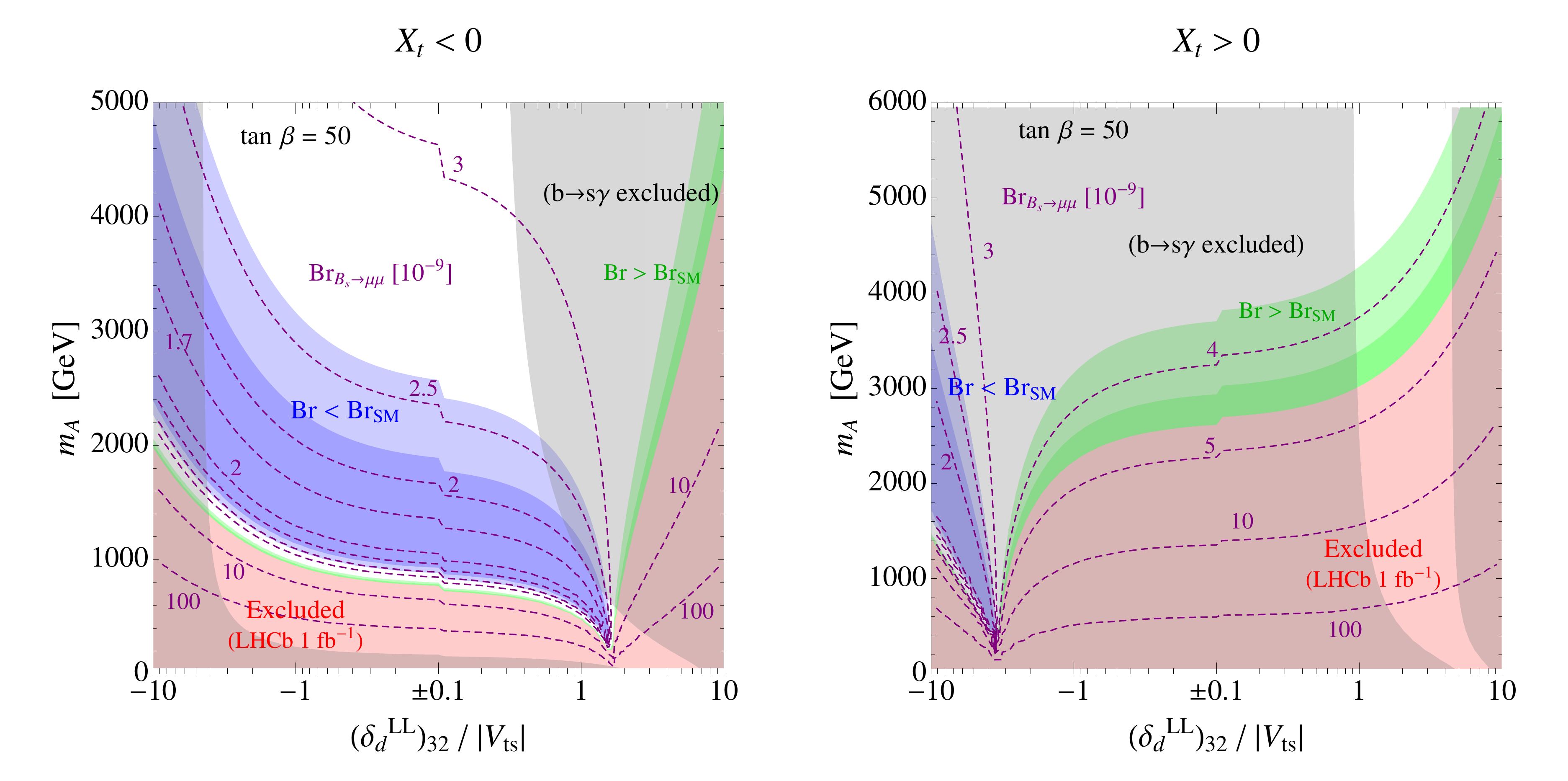}  \end{center}
\caption{\label{fig:BsMuMureach}
The reach for discovering a modification to $B_s \rightarrow \mu^+ \mu^-$, for the two points in parameter space denoted by black dots on figure~\ref{fig:bsg2D}, as a function of the heavy pseudoscalar Higgs mass, $m_A$, and  ${(\delta_d^{LL})}_{32} $ in units of $|V_{ts}|$.  The shaded gray region is excluded, at $2 \sigma$, by $b \rightarrow s \gamma$.  The purple contours indicate the branching ratio of $B_s \rightarrow \mu^+ \mu^-$ in parts per billion.  The red region is excluded at $2\sigma$ by LHCb with 1~fb$^{-1}$, $\mathrm{Br}_{B_s \rightarrow \mu^+ \mu^-} < 4.5 \times 10^{-9}$.  The blue (green) regions have a branching ratio that is smaller (larger) than the predicted SM cross-section of $(3.2 \pm 0.2) \times 10^{-9}$~\cite{Buras:2010mh} by $3\sigma$ in the region of light shading and $5 \sigma$ in the region with darker shading.
}
\end{figure}

The current limit, and reach, for discovering $B_s \rightarrow \mu^+ \mu^-$ are displayed in figure~\ref{fig:BsMuMureach}, as a function of the squark mixing, ${(\delta_d^{LL})}_{32}$, and the pseudoscalar Higgs mass, $m_A$, for the two points denoted by black dots in figure~\ref{fig:bsg2D}.  The shaded gray regions are excluded by $b \rightarrow s \gamma$ (note that we have included the charged Higgs contribution, which effects the boundary of the excluded region for $m_A \lesssim 1$~TeV).  The shaded red area is excluded by the LHCb limit on $\mathrm{Br}_{B_s \rightarrow \mu^+ \mu^-}$.  In nearly the entire parameter space this limit is significantly stronger than the direct LHC limit on heavy Higgses.  The strongest direct limit is set by CMS on $H \rightarrow \tau^+ \tau^-$, requiring $m_A \gtrsim 480$~GeV when $\tan \beta = 50$~\cite{Chatrchyan:2012vp}.  The (dark) blue region denotes values of the branching ratio of $B_s \rightarrow \mu^+ \mu^-$ $3 \sigma$ ($5 \sigma$) lower than the SM prediction\footnote{Note that the $3 \sigma$ and $5 \sigma$ regions of figure~\ref{fig:BsMuMureach} utilize the theory uncertainty of Ref.~\cite{Buras:2010mh}, where the $F_{B_s}$ uncertainty has been removed by assuming that the measurement of $\Delta M_{B_S}$ is SM-like.  Any deviation from the SM prediction needs to be treated with care because new physics can contribute to $\Delta M_{B_S}$.  In SUSY theories with MFV, new contributions to $\Delta M_{B_S}$ can be neglected, relative to new contributions to $B_s \rightarrow \mu^+ \mu^-$, because they rely on R-R squark mixing, which is small.}.
We see that a large range of $m_A$ leads to an observable depletion for the point with $X_t < 0$.  The green regions denote values that are enhanced by $3\sigma$ and $5 \sigma$.  This reflects a much smaller portion of the available parameter space because of the proximity of the stringent LHCb limit to the SM prediction.  

\begin{figure}[t]
\begin{center} \includegraphics[width=1\textwidth]{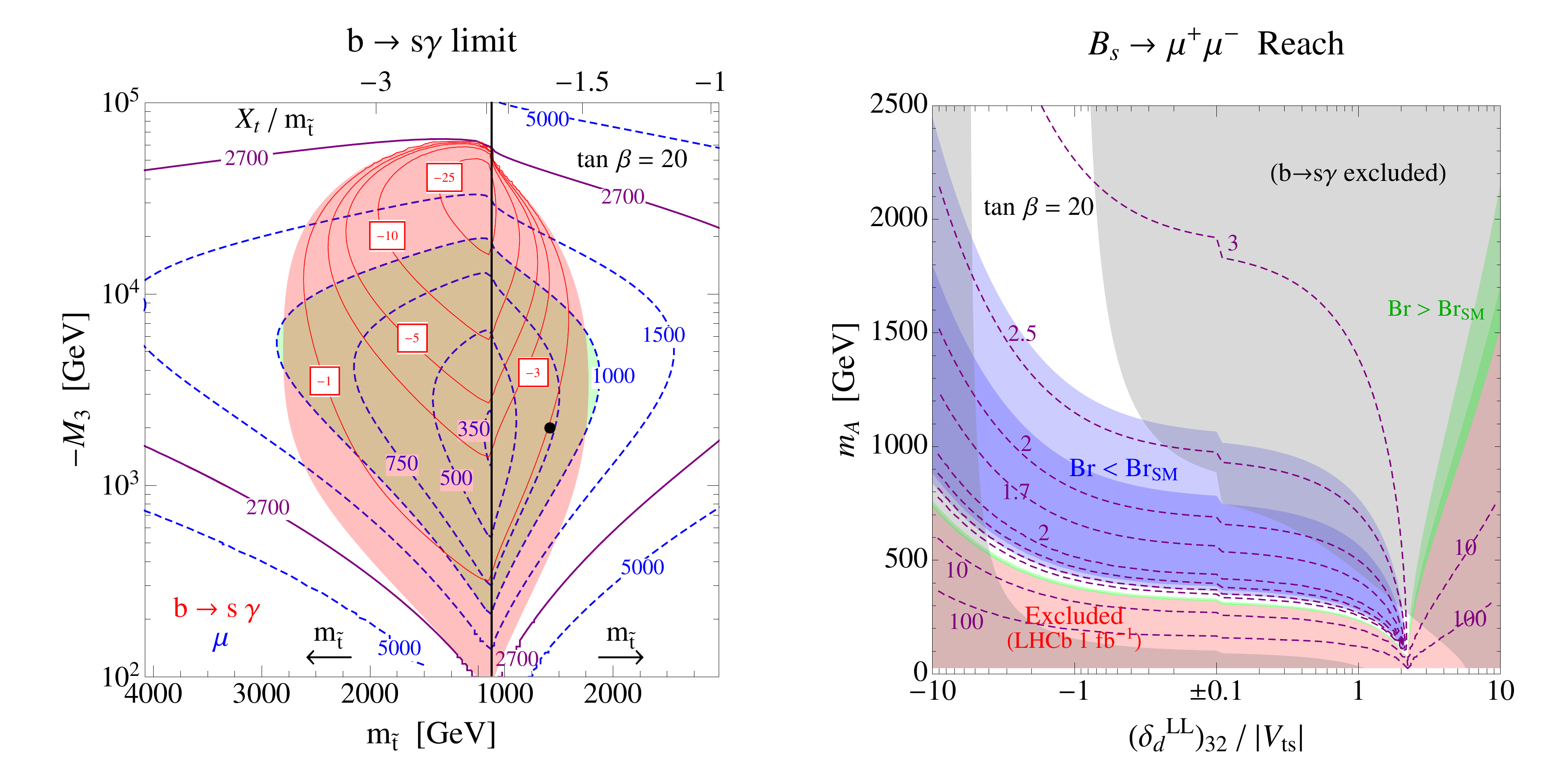}  \end{center}
\caption{\label{fig:flavorLowTb}
The limit on $b \rightarrow s \gamma$ and the reach for $B_s \rightarrow \mu^+ \mu^-$ are shown for $\tan \beta = 20$ on the left, and right, respectively.  The left plot corresponds to the region with $X_t < 0$ and all contours are as in figure~\ref{fig:bsg}.  The right plot corresponds to the point shown as a black dot in the left plot, and all contours and shading are as in figure~\ref{fig:BsMuMureach}.
}
\end{figure}

So far we have assumed $\tan \beta = 50$ in this section, as is appropriate for $t-b-\tau$ unification. 
But we found that $b-\tau$ unification is possible as long as $\tan \beta > 10$, and one may wonder how lowering $\tan \beta$ impacts the limits from $B$-meson decays.  In order to explore this, we choose the example value of $\tan \beta = 20$, and in figure~\ref{fig:flavorLowTb} we show the $b \rightarrow s \gamma$ limit on the left, for $X_t < 0$. Interestingly, we find that the $b \rightarrow s \gamma$ limit is stronger at lower values of $\tan \beta$: a larger fraction of the region with $\mu < 1$~TeV is excluded by the Higgsino-mediated diagram and therefore requires some cancellation between the Higgsino and gluino diagrams.  This might be seem surprising since the amplitude for $b \rightarrow s \gamma$ is proportional to $\tan \beta$.  However, so is the finite threshold to the bottom mass, and the $b \rightarrow s \gamma$ amplitude scales roughly as $\delta_b^{fin} / m_{\tilde t, \tilde b}^2$.   For a given value of $\mu$, as $\tan \beta$ is reduced the stop and sbottom masses must be lowered to keep $\delta_b^{fin}$ large enough for $b-\tau$ unification, resulting in a stronger $b \rightarrow s \gamma$ limit.  In addition, as we lower $\tan \beta$, a slightly larger $\delta_b^{fin}$ is required (see figure~\ref{fig:ModelIndep_btau}). Still, we find that only a mild cancellation is required between the Higgsino and gluino diagrams to bring $b \rightarrow s \gamma$ in accord with the limit, for most of parameter space, even as $\tan \beta$ is lowered.  

We now turn to the $B_s \rightarrow \mu^+ \mu^-$  reach at lower $\tan \beta$, which is shown to the right of figure~\ref{fig:flavorLowTb} for $\tan \beta = 20$ and the point in parameter space denoted by a black dot on the left side of the figure.  We see that the LHCb limit, and region with a reach to observe an enhanced or depleted branching ratio, are both pushed to lower values of $m_A$ relative to the $\tan \beta = 50$ scenario.  This is because the amplitude for $B_s \rightarrow \mu^+ \mu^-$ scales as $\tan^2 \beta \times \delta_b^{fin} / m_A^2$ (see equation~\ref{eq:BsMuMuAmp}), and the lower value of $\tan \beta$ is compensated by a lower value for $m_A$.  Note that a large fraction of the allowed values of $m_A$ still lead to an observable effect since the present limit is also reduced\footnote{For $\tan \beta = 20$, the LHC limit on the heavy Higgs is $m_A \gtrsim 300$~GeV~\cite{Chatrchyan:2012vp}, which is typically weaker than the LHCb limit on $B\rightarrow \mu^+ \mu^-$, as can be seen to the right of figure~\ref{fig:flavorLowTb}.}.

We now discuss the relationship between Yukawa Unification and the SUSY CP problem.  As is well-known, supersymmetry with generic CP phases is highly constrained by experimental limits on the neutron and electron Electric Dipole Moments (EDM).  In our parameter space, the sfermions have multi-TeV masses and this alleviates the SUSY CP problem relative to realizations of SUSY with lighter sparticles.  However, as we will see, constraints from EDMs are still relevant.

First we consider the neutron EDM\@. Because the leading contribution to the neutron EDM is generated by the same diagram as the left of figure~\ref{fig:bthresh_feyn} (except with an external photon and the bottom quarks switched to down quarks), there is a close relationship between the size of the neutron EDM and the gluino contribution to the finite $y_b$ threshold correction, $\delta_b^{\tilde{g}}$.
We note that the phase from the A-term is less constrained since diagrams analogous to the right diagram of figure~\ref{fig:bthresh_feyn} are suppressed by small CKM elements. For approximately degenerate squarks we can write the relation between the gluino contribution to the finite threshold and the neutron EDM~ \cite{Masiero} as
\be
d_n^{\tilde{g}} = \frac{2 e}{3} \textrm{Arg}(\mu M_3) \left(-\frac{g_3^2}{12 \pi^2} \frac{\mu M_3}{m_{\tilde{b}}^2} \textrm{tan} \beta\right) \frac{M_d}{m_{\tilde{d}}^2} \left( \frac{m_{\tilde{b}}}{m_{\tilde{d}}} \right)^2.
\label{eq:nEDM}
\ee
Since the neutron EDM is bounded by $d_n \leq 2.9 \times 10^{-26} \textrm{e cm} $ \cite{pdg} the bound on the CP violating phase goes as:
\be
\textrm{Arg}(\mu M_3) &\lesssim& 0.3 \left( \frac{0.1}{\delta_b^{\tilde{g}}} \right) \left( \frac{m_{\tilde{d}}}{3 \textrm{TeV}}\right)^2 \left( \frac{m_{\tilde{d}}} {m_{\tilde{b}}} \right)^2,
\label{eq:ArgmuBound}
\ee
where the last factor corrects for sbottom-sdown splitting.  We see that a generic phase is allowed for multi-TeV squarks and $\delta_b^{fin} \sim 0.1$, as is required for successful $b/\tau$ unification.

The experimental limit on the electron EDM is stronger, $d_e \leq 1.05 \times 10^{-27} \textrm{e cm}$~\cite{pdg}, and is generally more constraining than the neutron EDM\@. The leading contribution to the electron EDM comes from a loop mediated by charged Higgsinos and winos~\cite{Masiero}, which lead to a bound on Arg($\mu M_2$),
\be
\textrm{Arg}(\mu M_2) &\lesssim& 0.02  \left( \frac{50}{\textrm{tan}\beta} \right) \left(\frac{m_{\tilde{\nu}}}{5~\mathrm{TeV}}\right)^2,
\label{eq:electronBound}
\ee
where, for simplicity, we have fixed $\mu = M_2 = 1$~TeV\@. We see that for $\tan\beta \sim 50$, as favored for $t-b-\tau$ unification, the bound is stringent.  The bound is alleviated for smaller values of $\tan \beta$ (recall from above that $b-\tau$ unification works for $\tan \beta \gtrsim 10$ weakening the bound on the phase to $\sim 0.1$), or with a heavier sneutrino.

\subsection{Dark Matter}
\label{subsec:dm}

We have seen that requiring Yukawa unification along with $\mu \lesssim 1$~TeV, in order to avoid overclosure by the thermal relic density of a neutralino LSP, places an upper bound on the superpartner mass scale.  In this section we consider the allowed parameter space and the prospects for direct detection of the resulting WIMP dark matter.  We shall assume that the relic abundance of the LSP is given by thermal freeze-out.  While the parameter space of interest for Yukawa unification does contain single-species WIMP dark matter, we allow for the more general possibility that the LSP makes up only a subdominant component of the dark matter.  Multi-component dark matter can be obtained, for example, by environmental selection, which requires large, roughly equal relic densities for both WIMPs and axions due to dangerous boundaries in both the dark matter relic density and the vacuum misalignment angle~\cite{Hall:2011jd}.  Throughout this section we use a recent lattice value for the strange quark content of the nucleon~\cite{Giedt:2009mr}, $f_s = 0.069$; some results are compared with a larger $f_s$ in appendix~\ref{app:welltemp}\@.

Figure~\ref{fig:relicdensity} shows the region of parameter space with $\Omega_{\tilde{N}_1} \leq \Omega_{cdm}$, computed using MicrOMEGAs~\cite{Belanger:2006is,Belanger:2008sj}, shaded in orange, with contours of relic density normalized to $\Omega_{cdm}=0.111$ (from Ref.~\cite{pdg}) in black.   To be definite we once again fix $M_{1,2}$ using the gaugino unification relation, $6 M_1 \approx 3 M_2 \approx M_3$.  Since the neutralino annihilation and elastic scattering cross-sections do not depend sensitively on $M_{2,3}$, this assumption will not qualitatively affect our results as long as we have $|M_1| < |M_{2,3}|$, so that the LSP is a bino-Higgsino mixture.  The relic density contours show that the neutralino LSP makes up an $\mathcal{O}(1)$ fraction of dark matter in a large portion of the allowed parameter space.

\begin{figure}[t]
\begin{center} \includegraphics[width=1\textwidth]{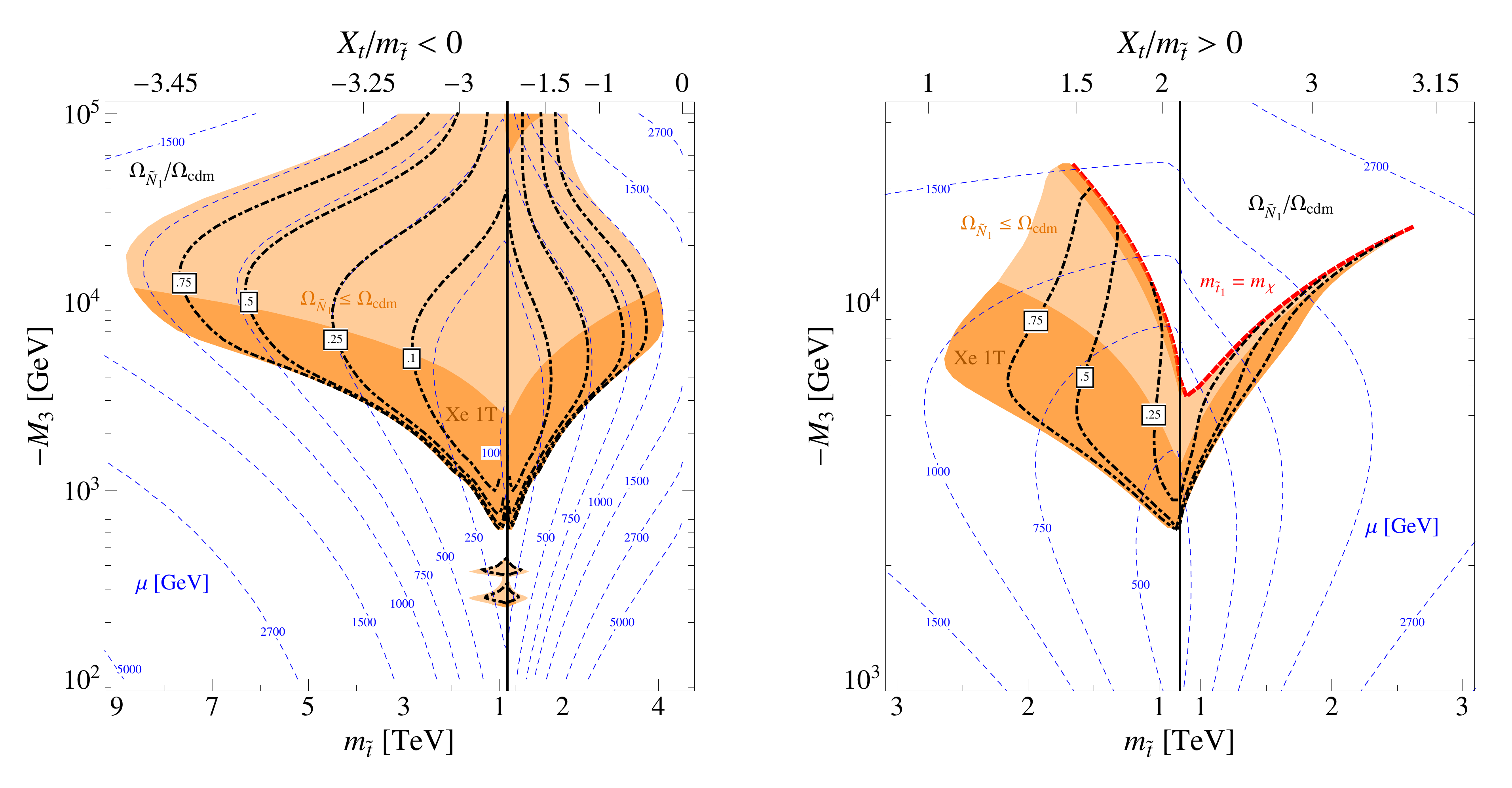}  \end{center}
\caption{\label{fig:relicdensity} The region of parameter space with $\Omega_{\tilde{N}_1} \leq \Omega_{cdm}$ are shown shaded in orange for $\tan\beta = 50$.  We have taken $m_A = 10$~TeV in order to remove the potential effects of the heavy Higgs funnel and imposed gaugino unification, $6 M_1 = 3 M_2 = M_3$.  The dashed blue contours show the value of the $\mu$-parameter required to achieve $b-\tau$ unification, while the black contours show the thermal relic density of the LSP normalized to the relic abundance of dark matter, $\Omega_{\tilde{N}_1}/\Omega_{cdm}$.  In calculating the relic abundance, we have taken all scalar masses degenerate at $m_{\tilde{t}}$ and fixed $A_t = A_b = A_\tau$ to the value required to obtain a 125~GeV Higgs.  All other $A$-terms are set to 0.  The darker orange shading depicts the predicted reach of the upcoming XENON1T direct detection experiment.}
\end{figure}

As in the previous sections, the dashed blue contours show the $\mu$-parameter that results from imposing $(t-) b-\tau$ unification and a 125~GeV Higgs.  In both panels of the figure, the lower boundary of the large orange regions corresponds to the ``well-tempered'' neutralino~\cite{ArkaniHamed:2006mb}. Moving to larger $|M_3|$, the LSP becomes dominantly Higgsino-like, and so the $\Omega_{\tilde{N}_1} = \Omega_{cdm}$ boundary tends to follow the $\mu \approx 1$~TeV contour, as expected for pure Higgsino dark matter.  As the stop mass decreases, however, Higgsino dark matter is required to be heavier on account of the increasing importance of stop- and sbottom-mediated annihilation channels whose amplitudes scale as $y_t^2$.  If the stop is lighter still, $m_{\tilde{t}} \gtrsim m_{\tilde{N}_1}$, then it may coannihilate with the LSP in a small, fine-tuned region, resulting in an even heavier neutralino.  

The LSP may also be dominantly bino-like, in the case where it annihilates resonantly through an $s-$channel mediator.  The Higgs and Z poles, respectively, are depicted in the left side of figure~\ref{fig:relicdensity} in the upper and lower islands below the large orange region.  Due to the assumption of gaugino unification, these regions contain a light gluino that is ruled out by LHC data; however, relaxing that assumption would allow for a heavier gluino.

\begin{figure}[t]
\begin{center} \includegraphics[width=1\textwidth]{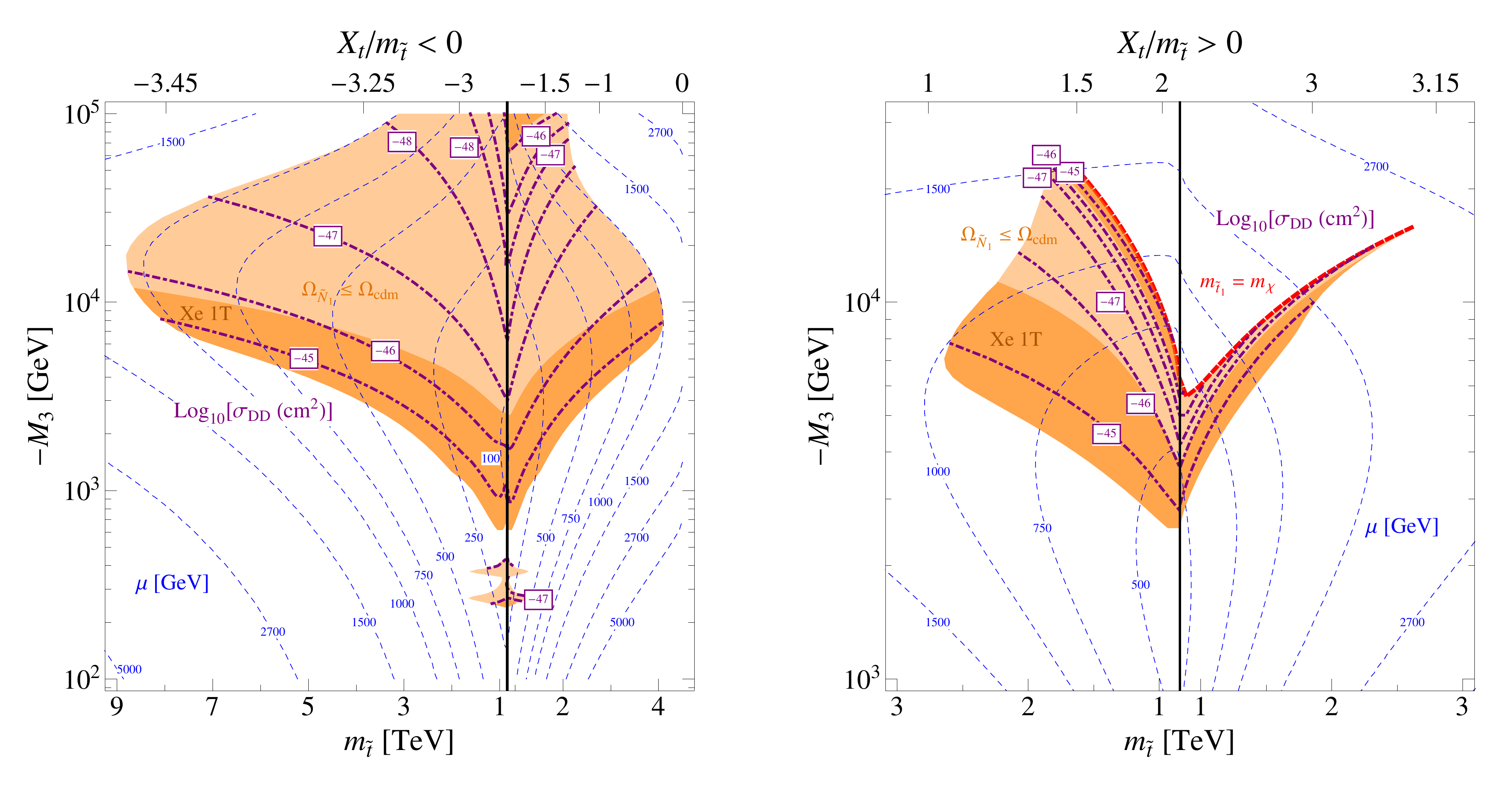}  \end{center}
\caption{\label{fig:directdetect} Contours of direct detection cross-section are shown in purple, with the orange shading and blue contours as in figure~\ref{fig:relicdensity}.  In terms of the elastic scattering cross-section off of protons, $\sigma_p$, we define the direct detection cross-section to be scaled by the LSP relic density, $\sigma_{DD} \equiv \sigma_p \left(\Omega_{\tilde{N}_1}/\Omega_{cdm}\right)$.  }
\end{figure}

The allowed parameter space will be further constrained by upcoming direct detection experiments, such as the planned ton-scale liquid xenon detector~\cite{Scovell}, whose predicted reach is shaded in figures~\ref{fig:relicdensity} and~\ref{fig:directdetect} in dark orange.  However, using the lattice prediction for the strange quark content of the nucleon~\cite{Giedt:2009mr}, there is no limit from current direct detection experiments.  Once again, we have used MicrOMEGAs to compute the direct detection cross-section.  Contours of the direct detection cross-section scaled by relic abundance, $\sigma_{DD} \equiv \sigma_p \left(\Omega_{\tilde{N}_1}/\Omega_{cdm}\right)$ are shown in purple in figure~\ref{fig:directdetect}, in which $\sigma_p$ is the cross-section for elastic scattering of the LSP off of a proton.

We see from figure~\ref{fig:directdetect} that XENON1T should cover the entire well-tempered region, up to and including $m_{\tilde{N}_1} = 1$~TeV, at which point the LSP becomes dominantly Higgsino-like.  For more detail on this portion of parameter space, including the expected reach of the upcoming XENON100 release, see appendix~\ref{app:welltemp}.  In this region the elastic scattering is dominated by Higgs exchange.  Increasing $|M_1|$ reduces the Higgs-exchange cross-section, since the LSP coupling to the Higgs is proportional to the bino content of the neutralino.  For pure Higgsino dark matter, the leading direct detection diagrams appear at one loop with a naive size of order $10^{-46}~\mathrm{cm}^2$~\cite{Cirelli:2005uq,Essig:2007az}; however, a few groups~\cite{Hisano:2011cs,Hill:2011be} find that an accidental cancellation between one and two-loop diagrams leads to a surprisingly small cross-section,  $\sigma_p \lesssim 10^{-48}~\mathrm{cm}^2$,  so that the tree-level Higgs exchange diagrams still dominate the one and two loop contributions over the entire parameter range of interest.

At sufficiently light squark masses, squark exchange diagrams with amplitudes scaling like $y_t^2$ can become important.  These diagrams destructively interfere with the Higgs exchange contributions, and at large enough bino mass, they can dominate the elastic scattering cross-section.  This effect accounts for the tendency of the direct detection cross-section to decrease and then increase again as the gaugino mass is increased with $m_{\tilde{t}} \sim 1-2$~TeV\@.  Here the direct detection cross-section is sufficiently small that this region of parameter space will not be probed by currently-planned experiments; however, UV considerations prefer a lighter gaugino mass.  RG running to low energies with a heavy gluino pulls up on the squark masses, while Yukawa unification and dark matter requirements bound the squark mass to be below $\sim 8.5$~TeV or $\sim 2.5$~TeV, depending on the sign of $X_t$.

Finally, we briefly comment on the effect of relaxing the assumption of gaugino unification.  If $M_1$ is made larger relative to $|M_3|$, then the well-tempered region moves to smaller gluino masses relative to the parameter space shown in figures~\ref{fig:relicdensity} and~\ref{fig:directdetect}, increasing the size of the parameter space that contains Yukawa unification without overclosure.  In this case LHC searches for the gluino may probe some or all of the well-tempered contour.  Furthermore, due to the shape of the $\mu$ contours, the well-tempered region would extend to lighter squark masses.  Deforming in the other direction, in which $M_1$ is taken lighter than $M_3/6$, the well-tempered contour moves to larger gluino masses.  This decreases the size of the parameter space, perhaps removing entirely the region with $X_t > 0$.  However, in this case direct detection experiments are able to probe a larger fraction of the overall space.  Thus we consider the direct detection prospects in the interesting regions of parameter space to be promising with the next generation of experiments.

\section*{Acknowledgments}

We thank Nima Arkani-Hamed, Michele Papucci, and Neal Weiner for useful discussions.  We also thank Pietro Slavich for providing us with the version of Suspect used in Ref.~\cite{Bernal:2007uv}, which computes the Higgs mass accurately with heavy scalars. This work was supported in part by the Director, Office of Science, Office 
of High Energy and Nuclear Physics, of the US Department of Energy under 
Contract DE-AC02-05CH11231 and by the National Science Foundation under 
grants PHY-0457315 and PHY-0855653.  J.T.R. is supported by a fellowship from the Miller Institute for Basic Research in Science.

\appendix

\section{Yukawa Unification with Extra Charged Matter}
\label{app:ExtraCharged}

Throughout this paper, we have restricted to the minimal field content of the MSSM and assumed that there is a desert between the superpartner mass scale and the scale of gauge coupling unification.  One might wonder how sensitive our results are to this assumption, since the presence of new states will modify the beta functions of the gauge couplings and Yukawas.  One possible addition to the field content that preserves the success of gauge coupling unification is the addition of complete GUT multiplets. In the context of an $SU(5)$ GUT, for example, there may be some number $N$ of extra $\mathbf{5}+\mathbf{\bar 5}$'s at an intermediate scale $M_{mess}$.  Extra charged messengers are typically included in models where supersymmetry is broken at a low-scale and mediated to the SM sector by gauge interactions.  Our main focus in this paper has been non-gravitino DM and not low-scale SUSY breaking.  Still, it is interesting to consider Yukawa unification in models with gauge mediation, and even if SUSY is broken at a high-scale, there may happen to be extra charged states present at intermediate scales.

\begin{figure}[h!]
\begin{center}  \includegraphics[width=0.5\textwidth]{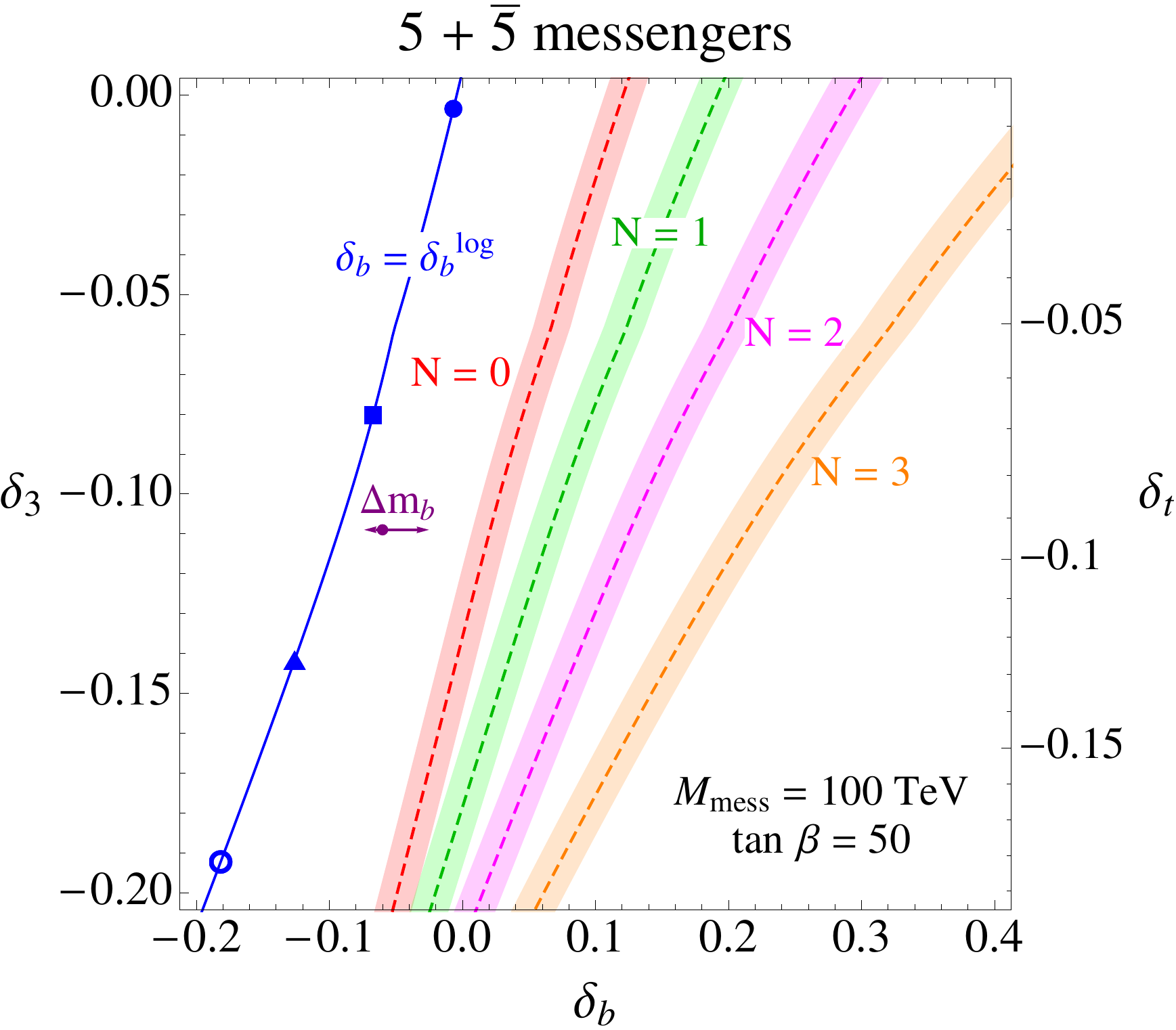} \end{center}
\caption{ \label{fig:ExtraCharged}
The required supersymmetric thresholds for $b-\tau$ unification in the presence of extra charged matter.  We fix $\tan \beta = 50$ and show the values of $\delta_b$ versus $\delta_3 / \delta_t$ necessary for precision $b-\tau$ unification with $N = 0,1,2,3$ extra $\mathbf{5}+\mathbf{\bar 5}$ with masses $M_{mess} = 100$~TeV\@.  The dashed lines indicate points with perfect unification, $\epsilon=0$, while the shaded regions denote $|\epsilon| < 0.02$.  The blue contour indicates $\delta_b = \delta_b^{fin}$, assuming degenerate superpartners except for $\mu = 500$~GeV\@.  The distances between the blue contour and the dashed lines indicate the necessary values of the finite threshold, $\delta_b^{fin}$, and we find that a larger finite threshold correction is required in the presence of extra charged states.
}
\end{figure}

In figure~\ref{fig:ExtraCharged} we show the necessary value of the $\delta_b$ and $\delta_3 / \delta_t$ thresholds, for precision $b-\tau$ unification, including $N= 0, 1, 2, 3$ extra $\mathbf{5}+\mathbf{\bar 5}$'s at $M_{mess} = 100$~TeV\@.  We choose $\tan \beta = 50$, as appropriate for $t -b-\tau$ unification, and for simplicity we set $\delta_{1,2}, \delta_\tau=0$ and $\delta_t = \delta_t^{log}$, as in figures~\ref{fig:ModelIndep_btau} and~\ref{fig:ModelIndep_tbtau}.   The blue curve indicates the value of $\delta_b^{log}$ that results assuming degenerate superpartners except for $\mu =$~500~GeV, as in the figures of section~\ref{sec:ModIndep}.  We see that a significantly larger value of the finite bottom threshold is required for $b-\tau$ unification in the presence of extra charged states.  This effect is easy to understand: the new fields increase the gauge coupling beta functions, leading to a larger value of $g_3$ that pulls down harder on the bottom Yukawa, making the precision of $b-\tau$ unification worse, before the finite bottom threshold is included.  If one imposes bino-Higgsino DM with $\mu < 1$~TeV, then the larger finite bottom threshold implies that the stop and sbottom must be even lighter than the minimal case with no extra charged states.

\section{Flavor Bases}
\label{app:flavor}
We define here the notation we use for flavor in section~\ref{subsec:Bmeson}.  We determined the flavor-dependent factors entering the amplitudes for $b \rightarrow s \gamma$ and $B_s \rightarrow \mu^+ \mu^-$ using the basis where the fermions are in mass eigenstate and the squark soft mass matrices are diagonal.  In the limit of small left-right mixing, this corresponds to the mass basis for the squarks and, as we review below, all flavor violation can be parameterized by the CKM matrix and the gluino vertices~\cite{Nir:1993mx}.  In the following, we specialize to the (s)quark sector, which is relevant for $B$-meson decays, and will not consider lepton flavor violation, which can be treated analogously.

As usual, the quarks are rotated from flavor to mass basis using unitary matrices $L_{u,d}$ and $R_{u,d}$,
\be
&u \rightarrow L_u \, u , \quad& d \rightarrow L_d \, d \nonumber \\
&u^c \rightarrow R_u \, u^c, \quad & d^c \rightarrow R_d \, d^c, 
\ee
where $V = L_u^\dagger L_d$ is the CKM matrix.  Meanwhile, we rotate the squarks to the basis where the soft masses, $m_{Q}^2, m_{U}^2, m_{D}^2$, are diagonal in flavor space,
\be
&\tilde u \rightarrow \tilde L_u \, \tilde u , \quad& \tilde d \rightarrow \tilde L_d \, \tilde d \nonumber \\
&\tilde u^c \rightarrow \tilde R_{u} \, \tilde u^c, \quad & \tilde d^c \rightarrow \tilde R_{d} \, \tilde d^c, 
\ee
where $\tilde L_u = \tilde L_d, \tilde R_u, \tilde R_d$ are unitary matrices.  The basis with diagonal squark soft masses is the mass eigenstate basis in the limit of small left-right mixing.  Note that left-right mixing is suppressed by the EW scale over the superparticle mass scale, $v / \tilde m$, and therefore this basis is approximately equivalent to the mass eigenstate basis in the parameter space relevant for Yukawa unification, $\tilde m \sim 1 - 10$~TeV\@.

In general, different rotations are needed to bring the quarks and the squarks to mass eigenstate, and SUSY flavor violation arises due to this misalignment.  After applying the above rotations,  flavor violation enters the vertices involving a quark, squark, and gaugino or Higgsino.  Flavor violation is encoded by the following unitary matrices,
\be
W^a = \tilde L_a^\dagger L_a \qquad W^{a^c}  =  \tilde R_a^\dagger R_a  \qquad a = u, d.
\ee
Note that $W^u$ and $W^d$ are related by a CKM rotation: ${W^u}^\dagger W^d = L_u^\dagger \tilde L_u \tilde L_d^\dagger L_d = L_u^\dagger L_d = V$.  The gluino vertices for $u$, $u^c$, $d$, $d^c$ are simply proportional to the corresponding $W^a$ matrices.  The charged Higgsino vertex, which also appears in section~\ref{subsec:Bmeson}, takes the form,
\be
Q \, y_u u^c \tilde H_u \supset \left( \tilde d \, W_d^* \, V^T \frac{m_u}{v \sin \beta} u^c + d \,  V^T \frac{m_u}{v \sin \beta} {W^{u^c}}^\dagger \tilde u^c \right) \tilde H_u^+, 
\ee
where $m_u$ is the diagonal up-type quark mass matrix.  The same expression describes the $\tilde H_d^-$ couplings, with the replacements $ u \leftrightarrow d$ and $\sin \beta \rightarrow \cos \beta$, $V^T \rightarrow V^*$.

Much of the literature on SUSY flavor uses a basis known as the SUSY-CKM basis, where the entire quark superfields are rotated into the fermionic mass basis by the $L_{u,d}$ and $R_{u,d}$ matrices.  In this basis, the squark soft masses are non-diagonal.  In the limit of small flavor violation, it is useful to work in the mass-insertion approximation,
\be \label{eq:mia}
\mathcal{M}_d^2 \, =  \tilde m_{d}^2 \, \,( \mathbb{1} + \delta_d) \qquad  \qquad \delta_d =
\left( \begin{matrix}
  \delta_d^{LL} & \delta_d^{LR} \\
  \delta_d^{RL} & \delta_d^{RR}
 \end{matrix} \right),
 \ee
and similarly for the up-type squarks.  The  SUSY-CKM basis is related to the basis described above, with diagonal squark soft masses, by applying rotations with the $W^a$ matrices, which diagonalize the $\delta_d^{LL}$ and $\delta_d^{RR}$ matrices,
 \be \label{eq:RelateBases}
 \mathrm{diag}(m_Q^2)  = \tilde m_{Q}^2 ( \mathbb{1} + W^d \, \delta_d^{LL} \, {W^d}^\dagger) \qquad \qquad   \mathrm{diag}(m_D^2)  = \tilde m_{D}^2 ( \mathbb{1} +  W^{d^c} \, \delta_d^{RR} \, {W^{d^c}}^\dagger ),
 \ee
 where the unit matrix in equation~\ref{eq:mia} results when $m_Q = m_D$.

 \section{Experimental Status of The Well-Tempered Neutralino}
\label{app:welltemp}

\begin{figure}[h!]
\begin{center}  \includegraphics[width=1\textwidth]{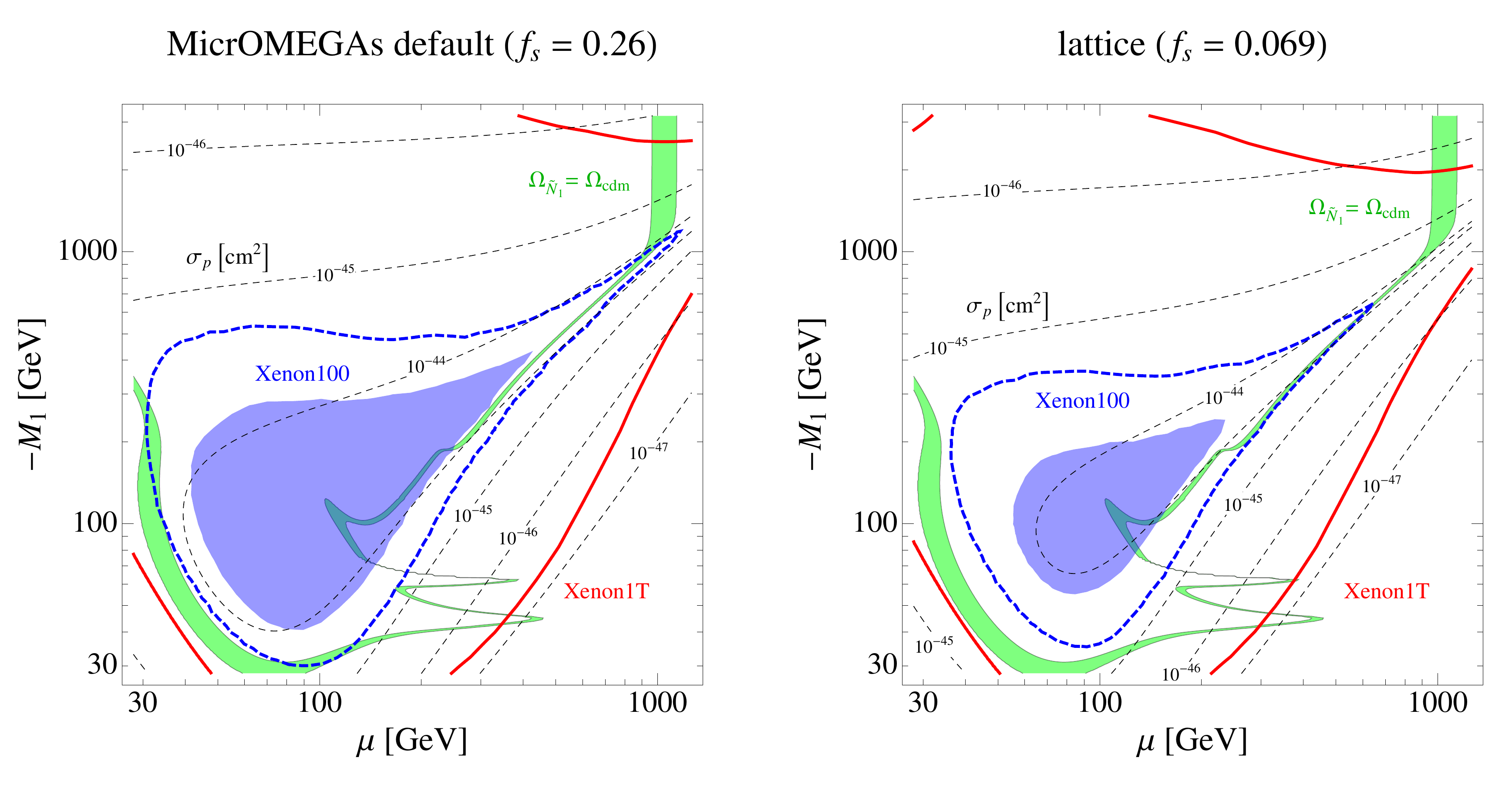} \end{center}
\caption{ \label{fig:welltemp}
The well-tempered parameter space in the $(\mu, M_1)$ plane, with $\tan\beta = 50$ and the strange quark content of the proton taken such that $f = 0.468$ and 0.319 for the left and right sides of the figure, respectively.  The green band corresponds to $\Omega_{\tilde{N}_1} = 0.111 \pm 0.018$, the 3$\sigma$ range for the relic abundance of dark matter, while the black dashed contours show the elastic scattering cross-section of the neutralino against a proton.  Note that, in contrast with the direct detection cross-sections plotted in figure~\ref{fig:directdetect}, $\sigma_p$ is not scaled by relic abundance.  The shaded blue region and dashed blue contour correspond to the current limits and projected reach of XENON100, while the red contour depicts the expected reach of the upcoming XENON1T experiment.  Here we decouple all scalar superpartners for simplicity, taking $\tilde{m} = m_A = 10$~TeV\@.}
\end{figure}

Here we consider in detail the well-tempered subset of the parameter space of section~\ref{subsec:dm}, in which the LSP is a bino-Higgsino mixture with the correct thermal relic abundance to be all of dark matter.  This corresponds to a portion of the boundary of the region in figures~\ref{fig:relicdensity} and~\ref{fig:directdetect}.  For simplicity we shall decouple all scalar superpartners, in which case the dominant elastic scattering diagram comes from Higgs exchange.  In this case, the elastic scattering cross-section is proportional to $|f|^2$, where
\be
f \equiv \frac{2}{9} + \frac{7}{9}\sum_{q = {u,d,s}} f_q \qquad \qquad f_q \equiv \frac{\langle N | m_q \bar{q} q | N \rangle}{m_N}.
\ee

Thus the ability of current and future direct detection experiments to probe the well-tempered region depends sensitively on the strange quark content of the nucleon, $f_s$.  Following the treatment in~\cite{Farina:2011bh}, figure~\ref{fig:welltemp} shows the well-tempered parameter space in the $(\mu, M_1)$ plane, with the $3\sigma$ range corresponding to the cosmological abundance of dark matter shaded in green.  In addition to the MicrOMEGAs default value of $f_s$, which gives $f = 0.468$~\cite{Belanger:2008sj}, we show the effect of taking the value of $f_s$ suggested by more recent lattice calculations, which gives $f = 0.319$~\cite{Giedt:2009mr}, considerably reducing the reach of direct detection experiments.  Furthermore, in addition to the current XENON100 limit~\cite{Aprile:2011hi}, shaded in blue, we show the predicted reach for both the imminent XENON100 data release, the dashed blue contour, as well as the upcoming XENON1T experiment~\cite{Scovell}, drawn in red.

Given the larger value of $f_s$, XENON100 excludes LSP masses between $m_h \lesssim m_{\tilde{N}_1} \lesssim m_t$ and will soon be able to constrain nearly the entire well-tempered parameter space, except for very light neutralinos and those which annihilate resonantly through a Z or Higgs boson.  However, if the lattice calculations are to be believed, then XENON100 only currently rules out a small region around $m_{\tilde{N}_1} \approx m_W$, and it will only be able to probe dark matter masses up to $\sim 600$~GeV\@.  In either case, however, XENON1T will probe the entire well-tempered parameter space up to $\mu \sim 1$~TeV, at which point the LSP becomes dominantly Higgsino-like.  Note that these bounds assume the best-case scenario: elastic scattering diagrams involving squark exchange interfere destructively with the Higgs exchange diagram, so that the bounds weaken as the squarks become lighter.

\end{document}